\documentclass[11pt]{article}
\usepackage{axodraw}
\usepackage{amsfonts}
\usepackage{epsfig}
\usepackage{amssymb}
\newcommand{\mrm}[1]{\mbox{\rm #1}}
\newcommand{\nn}{\nonumber}
\newcommand{\be}{\begin{equation}}
\newcommand{\bea}{\begin{eqnarray}}
\newcommand{\eea}{\end{eqnarray}}
\newcommand{\ee}{\end{equation}}

\newcommand{\eq}[1]{Eq.~(\ref{#1})}
\newcommand{\D}{D\hspace{-8pt}\slash}
\newcommand{\dv}{\partial\hspace{-7pt}\slash}

\newcommand{\dvr}{\stackrel{\hspace{3pt}\rightarrow}{\partial\hspace{-7pt}\slash}}

\newcommand{\gev}{ {\rm GeV} }

\begin{document}
\thispagestyle{empty}
\begin{flushright}
{\tt hep-ph/0307058}\\
{FTUAM-03-12}\\
{IFT-UAM/CSIC-03-26} \\
{UCSD/PTH 03-08}

\end{flushright}
\vspace*{1cm}
\begin{center}
{\Large{\bf  Neutrino Physics in the Seesaw Model } }\\
\vspace{.5cm}
A. Broncano$^{\rm a,}$\footnote{alicia.broncano@uam.es}, 
M.B. Gavela$^{\rm a,}$\footnote{gavela@delta.ft.uam.es} and 
E. Jenkins$^{\rm b,}$\footnote{ejenkins@ucsd.edu}
 
\vspace*{1cm}
$^{\rm a}$ Dept. de F\'{\i}sica Te\'orica, C-XI, and 
IFT, C-XVI, Facultad de Ciencias, \\ 
Univ. Aut\'onoma de Madrid, Cantoblanco, 28049 Madrid, Spain \\
$^{\rm b}$ Dept. of Physics, University of California at San Diego, 
9500 Gilman Drive, La Jolla, CA 92093, USA

\vspace{.3cm}

%
%
\begin{abstract}

The seesaw model of heavy and light Majorana neutrinos and its 
low-energy effective theory are studied, when the number of 
heavy neutrinos is equal to or less than the number of light 
lepton generations.  We establish a general relationship between the
high-energy parameters and the low-energy observables involving only 
the light fields.  It is shown how low-energy measurements of the properties 
of light neutrinos suffice {\it a priori} to determine all couplings of the unobserved heavy neutrinos.  $CP$ violation is present in low-energy processes 
if seesaw-model leptogenesis creates the matter-antimatter 
asymmetry of the universe.

\end{abstract}

\end{center}
%
%

\pagestyle{plain} 
\setcounter{page}{1}
\setcounter{footnote}{0}
\setcounter{section}{0}

\section{Introduction}

The leading high-energy theory of neutrino mass is the seesaw model \cite{seesaw}, 
the minimal extension of the Standard Model (SM) which includes 
right-handed neutrino singlets with heavy Majorana masses.  The seesaw model
has a number of attractive features: 
(1) It explains
the lightness of weakly-interacting neutrinos and the heaviness of sterile
neutrinos via the seesaw mechanism~\cite{seesaw}. 
(2) It provides a natural solution for the generation of the matter-antimatter
asymmetry of the universe. Leptogenesis occurs at high energies 
in heavy neutrino decay~\cite{fy}.  Approximately half of this lepton asymmetry
is converted into a baryon asymmetry.
(3) It is a renormalizable high-energy theory, so the number of high-energy
parameters is finite.
(4) It is natural: once right-handed
neutrinos are added to the SM matter content, the Lagrangian
containing all renormalizable terms allowed by 
$SU(3) \times SU(2)_L \times U(1)_Y$ 
gauge symmetry is the seesaw Lagrangian.
(5) It can be embedded into a unified or partially unified theory at a higher
energy scale.  The inclusion of right-handed neutrinos is natural in theories
with extended gauge interactions at high energy, such as $SU(4) \times SU(2)_L
\times SU(2)_R$, $SU(5) \times U(1)$, $SO(10)$ and $E6$.  In theories of 
this type, the asymmetry of the
gauge interactions of left-handed and right-handed fermions, as well as the
asymmetry between the heavy and light Majorana neutrinos, is a consequence of
the symmetry breakdown of the vacuum state rather than an asymmetry in the
underlying gauge interactions.  The Majorana mass term for the right-handed
neutrinos is the order parameter for breaking the unified or partially unified
gauge symmetry down to $SU(3) \times SU(2)_L \times U(1)_Y$.

An alternative theory for neutrino mass
is obtained by making the \emph{ad hoc} assumption that lepton number
$L$ is exactly conserved.  
In this theory, neutrinos acquire
only Dirac masses via the standard Higgs mechanism of electroweak symmetry
breaking since $L$-violating Majorana neutrino masses are forbidden.
The model of Dirac neutrino mass is disfavored relative to the seesaw
model for the following reasons: 
(1) It fails to explain the mass hierarchy between neutrinos and other
fermions.  In order to account for experimental data,
the Yukawa couplings of neutrinos must be many orders of magnitude smaller
than those of all the other fermions.
(2) The exact conservation of $L$, and consequently $(B-L)$ so long as baryon
number $B$ is conserved, forbids 
generation of the baryon asymmetry at energy scales above the
electroweak scale.
(3) It is unnatural, since Majorana mass terms of right-handed neutrinos are
not forbidden by the SM gauge symmetry. 

Although very light neutrino masses become natural in the seesaw model,
the electroweak hierarchy problem remains, since there are radiative corrections 
to the Higgs mass which are quadratic in the new heavy Majorana neutrino scale.
As is well-known, supersymmetry could soften this 
electroweak fine-tuning problem.
We concentrate, however, on the minimal (non-supersymmetric) 
seesaw model throughout this paper,
in order to study the essential properties of the seesaw mechanism.
Any beyond-the-SM theory, supersymmetric or not, which incorporates 
the seesaw mechanism, naturally  
is expected to include the general properties analyzed here.

In order to establish whether the seesaw model is the theory of nature,
it is important to work out the low-energy implications of the seesaw theory.
Our aim is to establish the connection of low-energy observables to the
parameters of the underlying high-energy minimal seesaw model, without any {\it ad hoc}
assumption on the values of the latter. In particular, it is very important 
to determine the connection of the $CP$-odd phases required for leptogenesis and 
a hypothetical future discovery of low-energy leptonic $CP$-violation. Our aim is to go 
farther than the usual general assertion that the low-energy phases are general 
combinations of all high-energy phases and moduli.

At low energies, there are two types of phases. Those contributing to 
$\Delta L=0$ transitions,
for instance to neutrino-neutrino oscillations (such as the leptonic $CKM$-like 
phase $\delta$ in the case 
of three light neutrino generations \cite{CKM}), are customarily called Dirac phases. 
Those contributing only to $\Delta L\ne 0$ transitions are designated 
Majorana phases \cite{Majorana1}.
Information on the latter could be extracted, in principle, from a positive 
measurement of neutrinoless double beta decay, combined with a determination of 
the absolute neutrino mass scale from tritium decay and with
the observed values of the leptonic mixing angles.
Even assuming that all those positive signals are indeed obtained in the future, 
a definite claim on the discovery of $CP$-violation from those data may be 
extremely difficult, if not impossible \cite{Glashow}, although the subject is 
debatable \cite{PPR1}.
On the contrary, the $CKM$-like leptonic phase $\delta$ may well be in reach 
in future superbeam facilities and/or neutrino factories \cite{SB,NF}, provided the  
leptonic angle $\theta_{13}$ is not too small.
A hypothetical measurement of $\delta$, together with the determination of the 
Majorana character of the neutrinos, would not be, by itself, a proof of 
leptogenesis as the origin of the matter-antimatter asymmetry of the universe, but 
it would make the argument for leptogenesis extremely compelling. 

In Ref.~\cite{bgj}, we constructed a low-energy effective theory for the 
seesaw model by integrating out the heavy Majorana neutrinos.  The effective
Lagrangian is of the form
\be\label{leff}
{\cal L}_{\rm eff} = {\cal L}_{\rm SM} + \delta{\cal L}^{d=5} + \delta{\cal
L}^{d=6} + \cdots,
\ee
where ${\cal L}_{\rm SM}$ is the Standard Model Lagrangian and the 
higher-dimensional 
operators give the low-energy physics effects of the heavy Majorana neutrinos.  

The leading higher-dimensional operator in the effective theory
is the unique $d=5$ operator~\cite{weinberg} which can be built with Standard Model fields:  
\be\label{d5}
\delta{\cal L}^{d=5} = \frac{1}{2}\, c_{\alpha \beta}^{d=5} \,
\left( \overline{{\ell_{L\alpha}}^c} \tilde \phi^* \right) \left(
\tilde \phi^\dagger \, \ell_{L \beta} \right) + {\rm h.c.},
\ee 
where $\ell_L$ are the lepton weak doublets and $\tilde \phi$ is related to
the standard Higgs doublet by $\tilde \phi = i \tau_{2} \phi^*$. 
The $d=5$ operator coefficients
are given in terms of the parameters of the
high-energy seesaw theory by 
\be\label{cd5}
c^{d=5} = \left( Y_\nu^* \, (M^*)^{-1} \, Y_\nu^\dagger \right)\ ,
\ee 
where $M$ denotes the heavy neutrino mass matrix, and $Y_\nu$ is
the Yukawa coupling matrix between the heavy neutrinos and the charged lepton doublets. 

When the electroweak gauge symmetry breaks
spontaneously, this $d=5$ operator yields a $d=3$ Majorana mass term for the
left-handed weakly interacting neutrinos.  Rewriting the theory in terms of
the Majorana mass eigenstate neutrinos leads to the presence of a non-trivial
lepton mixing matrix in the couplings of $W$ bosons to lepton charged currents. 

In Ref.~\cite{bgj}, the $d=6$ term of the low-energy effective theory was determined 
to be
\be\label{d6}
\delta{\cal L}^{d=6} = c^{d=6}_{\alpha \beta} \, \left( \overline{\ell_{L\alpha}} \tilde \phi
\right) i \dv \left( \tilde \phi^\dagger \ell_{L \beta} \right),
\ee 
where the $d=6$ operator coefficients are given in terms of the parameters
of the high-energy seesaw theory by
\be\label{cd6}
c^{d=6} = Y_\nu \, (|M|^2)^{-1} \, Y_\nu^\dagger \ .
\ee  
When the Higgs doublet acquires a vacuum
expectation value, this $d=6$ operator leads to $d=4$ kinetic energy terms for
the left-handed Majorana neutrinos. 
Other $d=6$ operators will be present in the low-energy 
Lagrangian, since they are generated by radiative mixing of the above operator
in the renormalization
group running between the high-energy and  
low-energy scales.  These other $d=6$ operators
will not be discussed in this work, as their effects are subdominant \cite{us}.

In Ref.~\cite{bgj}, we showed that the low-energy
effective theory including only the $d=5$ and $d=6$ operators contains an equal 
(a greater) number of real and imaginary parameters as the high-energy 
seesaw model when the number of right-handed neutrinos in the seesaw theory 
is equal to (less than) the number of 
generations of Standard Model fermions.  Thus, low-energy measurement of the 
$d=5$ and $d=6$ operator coefficients suffices {\it a priori} to determine all of 
the parameters of the high-energy seesaw theory.

While the direct physical effects of the 
$d=5$ operator, such as the light neutrino masses, are proportional to $1/M$, 
those of the $d=6$ operator are suppressed by $1/M^2$ and their experimental 
measurement is presently out of reach, at least for natural values of the 
heavy seesaw scale. 
Nevertheless, the ensemble of the  $d=5$ and $d=6$ operators is the low-energy 
tell-tale of the seesaw mechanism for any theory beyond the SM which includes 
that mechanism, barring most unnatural cancellations, and it deserves particular study.
In this work, we make explicit the connection of the high-energy parameters of the 
seesaw model to the $d=5$ and $d=6$ operator coefficients. We will see that the analysis
of this operator structure sheds light on the relation between the seesaw model 
and planned experiments.

We work out the general relation between the high-energy seesaw parameters 
in $Y_\nu$ and $M$ and the $d=5$ and $d=6$ operator coefficients of the
low-energy effective theory, for an arbitrary number 
of light generations. This analysis allows us to express the combination of 
high-energy phases contributing to leptogenesis in terms of only low-energy
parameters in a very compact form.  The most important physical consequences 
will be already transparent from the general 
formulae, which are basis-independent.

For detailed computations, it is necessary to specify a basis.  A standard 
high-energy basis for the seesaw model is defined.  No unphysical parameters 
occur in the standard high-energy basis; for this reason, it is the simplest 
choice of basis.  Besides including the customary choice of diagonal $M$ and 
$Y_e$, the standard high-energy basis expresses the seesaw theory in terms of 
physical, basis-independent, moduli and phases of $Y_\nu$.  For the case of 
two generations, $Y_\nu$ contains 2 high-energy phases, whereas for three 
generations, it contains 6 phases.  The connection between the standard 
high-energy basis and the low-energy bases is made.

The lessons learned in the general case will be illustrated in full detail for 
the minimal seesaw model with two heavy and light generations, as an important 
special case.  The generalization to three generations is straightforward, 
but more complicated, and will be published in a separate work.  Leptogenesis 
already occurs for two generations of leptons since there are $CP$-odd phases 
present in the lepton sector due to the Majorana character of neutrinos.  Thus, 
interesting $CP$-violating effects occur in low-energy processes with only two 
generations of leptons, and can be studied in this simpler context.  The low-energy 
effective theory including the $d=5$ operator contains only a single $CP$-odd phase 
in the lepton mixing matrix, while there are two independent phases at high energies.  
Thus, not all of the parameters in the seesaw model appear in 
the low-energy effective theory with only the $d=5$ operator.   
Indeed, the $d=6$ operator coefficient contains an additional
$CP$-violating phase which is independent of the phase of 
the lepton mixing matrix. It is very instructive to 
work out in detail the connection between leptogenesis and the 
two low-energy $CP$-odd phases.

The organization of the paper is as follows. 
 Section 2 reviews the 
high-energy seesaw Lagrangian.  The standard high-energy basis for 
the seesaw model is defined, and the leptogenesis asymmetry 
is discussed.  Section 3 presents the $d\le6$-effective Lagrangian in 
the low-energy spontaneously-broken theory.  Physical observables and 
global symmetries of the effective Lagrangian are discussed as well.  
Section 4 gives the general 
connection between the high-energy parameters of the seesaw model and 
the low-energy parameters of the effective theory, and relates high-energy 
and low-energy observables. 
The relationship of our results with those in
the literature also is discussed.  Section 5 specializes the previous general 
results to the seesaw model with two generations of heavy neutrinos and 
light leptons.  Additional results are provided in the appendices.  
Appendix 1 considers the relationship of our standard high-energy basis 
to alternative high-energy bases with additional unphysical parameters, 
for any number of lepton generations.  Appendices 2, 3 and 4 concern the case 
of two generations: they contain, respectively, formulae relating the flavor and mass 
eigenstate bases of the light neutrinos, the generalizations of 
the low-energy observables to include the effects of the $d=6$ operator, and
the determination of the heavy Majorana neutrino masses from low-energy observables.
Appendix 5 contains the Feynman rules for neutrinos at 
high and low energies.

\section{Seesaw Lagrangian}

The seesaw Lagrangian is the most general renormalizable Lagrangian which
can be written for the Standard Model gauge group with right-handed neutrinos
included in the fermion content of the theory.
The leptonic Lagrangian of the seesaw model 
is given by:
\be
\label{SS_lagr}
{\cal L}_{\rm leptons}= {\cal L}^{\rm KE}_{\rm leptons}
+{\cal L}^{\chi \rm SB}_{\rm leptons},
\ee
where 
\be
{\cal L}^{\rm KE}_{\rm leptons} = i \, \overline{\ell_L} \,
\D \, \ell_L + i \, \overline{e_R} \, \D \, e_R
+ i \, \overline{N_R} \, \dv \, N_R  \\
\ee
contains the kinetic energy and gauge interaction terms 
of the left-handed lepton doublets $\ell_L$, 
the right-handed charged leptons $e_R$, and
the right-handed neutrinos $N_R$, and
\be
{\cal L}^{\chi \rm SB}_{\rm leptons} =
-  \overline{\ell_L} \,{\phi} \, {Y_e}  \, e_R
- \overline{\ell_L} \,{\widetilde\phi} \, {Y_\nu}  \, N_R
-\frac{1}{2}\,\overline{{N_R}^c} \, M \,{N_R}
+ \mrm{h.c.}\,,
\ee
contains the Yukawa interactions and the Majorana mass term of 
the gauge-singlet right-handed neutrinos.  
The kinetic energy Lagrangian of the lepton sector is invariant under
the chiral transformations
\be\label{chiral1}
\ell_L \to V_\ell \, \ell_L \,, \quad e_R \to V_e \, e_R \,, 
\quad N_R \to V_N \, N_R \,,
\ee
where $V_\ell$ and $V_e$ are $n \times n$ unitary matrices for
$n$ generations of light leptons, and $V_N$ is an $n^\prime \times n^\prime$ unitary
matrix for $n^\prime$ generations of right-handed neutrinos.
The chiral symmetry $G= U(n)_\ell \times U(n)_e \times U(n^\prime)_N$ of the lepton
kinetic
energy terms is completely broken by the chiral symmetry-breaking Lagrangian
${\cal L}^{\chi \rm SB}$.  ${\cal L}^{\chi \rm SB}$ would respect
the chiral symmetry if the matrices $Y_e$, $Y_\nu$ and $M$ transformed as 
\bea
\label{chiral}
Y_e &\to& V_\ell \, Y_e \, V_e^\dagger\,, \nn\\
Y_\nu &\to& V_\ell \, Y_\nu \, V_N^\dagger\,, \\
M &\to& V_N^* \, M \, V_N^\dagger\,, \nn
\eea
under the chiral symmetry.  Eq.~(\ref{chiral}) implies that
seesaw theories with Yukawa
and Majorana mass matrices related by chiral transformations are equivalent
theories.  Thus, it is possible to use the chiral transformations 
to write the explicit
chiral symmetry-breaking parameters in a convenient basis.  As shown in
Ref.~\cite{bgj}, the number of physical parameters 
in $Y_e$, $Y_\nu$ and $M$ is equal to the number of parameters in
$Y_e$, $Y_\nu$ and $M$ minus the number of parameters in $V_\ell$, $V_e$
and $V_N$.  The seesaw model contains $(n+n^\prime+nn^\prime)$ real 
and $n(n^\prime-1)$ imaginary 
physical parameters in the matrices $Y_e$, $Y_\nu$, and $M$.

Let us use the chiral transformations of the high-energy seesaw model to define
a standard high-energy basis which does not contain any spurious unphysical
parameters. It will include the convenient and customary specialization to a basis in
which both $Y_e$ and $M$ are diagonal and real.

In a general basis, the complex $n \times n$ matrix $Y_e$ depends on $n^2$ real
and $n^2$ imaginary parameters. Let us denote by $y_\alpha$ the eigenvalues of $Y_e$, 
with $\alpha$ being the flavor index
which, in a redundant notation, takes the values $\alpha=1,2,\cdots, n= e, \mu, \cdots, e_n $.
 The change of basis
\be
Y_e \to V_\ell \, Y_e \, V_e^\dagger = \left( Y_e \right)_{\rm diag, \, real}
= {\rm diag}\left( y_e,y_\mu, \cdots, y_{e_n} \right),
\ee
determines $n(n-1)$ real and $n^2$ imaginary parameters of the matrices
$V_e$ and $V_\ell$, which each contain $n(n-1)/2$ real and $n(n+1)/2$ imaginary
parameters.  All of the real and $n^2$ of the $n(n+1)$ imaginary parameters
in $V_e$ and $V_\ell$ are determined by diagonalization of $Y_e$;
 $n$ imaginary parameters in $V_e$ and $V_\ell$ are not determined 
since $\left( Y_e \right)_{\rm diag, \, real}$ is left invariant by a transformation with
\be
\label{V_ell}
V_\ell = V_e = {\rm diag}\left(e^{i\zeta_1}, e^{i\zeta_2}, \cdots , e^{i\zeta_n} \right)\,,
\ee
which depends on $n$ arbitrary phases $\zeta_\alpha$, $\alpha = 1, \cdots, n$.

In a general basis, the complex symmetric $n^\prime \times n^\prime$ matrix $M$ contains
$n^\prime(n^\prime+1)/2$ real and $n^\prime(n^\prime+1)/2$ imaginary parameters.  The change of basis
\be\label{Mdiag}
M \to V_N^* \, M \, V_N^\dagger = M_{\rm diag, \, real} = 
{\rm diag}\left( |M_1|, |M_2|, \cdots , |M_{n^\prime}| \right),
\ee
fully determines the $n^\prime(n^\prime-1)/2$ real and $n^\prime(n^\prime+1)/2$ imaginary parameters of the
$n^\prime \times n^\prime$ unitary matrix $V_N$.  $M_{\rm diag, \, real}$ depends on the $n^\prime$
real mass eigenvalues of the heavy neutrinos.  No further transformation by $V_N$ is
allowed.  In this basis, the heavy Majorana
mass eigenstate neutrinos are
\be
N_i = N_{Ri} + {N_{Ri}}^c \, ,
\ee
and satisfy the condition ${N_i}^c = N_i$.

The leptonic Lagrangian of the seesaw model, expressed in terms of the heavy
Majorana mass eigenstate neutrinos $N_i$, is then given by Eq.~(\ref{SS_lagr}) with
\bea
{\cal L}^{\rm KE}_{\rm leptons} &=& i \, \overline{\ell_L} \,
\D \, \ell_L + i \, \overline{e_R} \, \D \, e_R
+ \frac{1}{2} i \, \overline{N}_i \, \dv \, N_i , \\
{\cal L}^{\chi \rm SB}_{\rm leptons} &=&
-  \overline{\ell_L} \,{\phi} \, {Y_e}  \, e_R
- \frac{1}{2}  \overline{\ell_L} \,{\widetilde\phi} \, 
{Y_\nu} \, N_i
-\frac{1}{2}\,\overline{N}_i \, |M_i| \,{N}_i
+ \mrm{h.c.}
\eea

In the basis in which $Y_e$ and $M$ are both diagonal and real, the complex
$n \times {n^\prime}$ matrix $Y_\nu$ contains $n n^\prime$ real
and $n n^\prime $ imaginary parameters, which are the magnitudes and phases
of the $Y_\nu$ matrix elements
\be
\label{Ynugen}
\left( Y_\nu \right)_{{\alpha} j} \equiv u_{{\alpha} j} \, e^{i\phi_{\alpha j}} \ ,
\ee
where $j = 1,\cdots, n^\prime$ is the heavy Majorana neutrino
mass eigenstate index.

Under a $V_\ell = V_e$ rephasing, $\phi_{\alpha j} \to \phi_{\alpha j} +\zeta_\alpha$.
To specify the phases $\zeta_\alpha$ is to choose a basis.  
The $n n^\prime$ phases in the matrix $Y_\nu$ consist of 
$n(n^\prime-1)$ independent phases,
\bea\label{offdiagphase}
\Phi_{\alpha j} &\equiv& \left( \phi_{\alpha j} - \Psi_\alpha \right)\, 
= \phi_{\alpha j}
- {1 \over n^\prime} \, \sum_{k =1}^{n^\prime} \, \phi_{\alpha k} \, , 
\eea
which are invariant under $\zeta_\alpha$ rephasings, and 
$n$ phases,
\be
\Psi_\alpha \equiv {1 \over n^\prime} \sum_{j=1}^{n^\prime} \, \phi_{\alpha j}\ ,
\ee
which transform under $\zeta_\alpha$
rephasings as $\Psi_\alpha \to \Psi_\alpha + \zeta_\alpha$.  The basis-dependent
phases $\Psi_\alpha$ are unphysical, and can be removed from $Y_\nu$ by fixing
\be
\label{zetahe}
\zeta_\alpha = -\Psi_\alpha\,.
\ee
The phases in each row of $Y_\nu$ then satisfy the constraint
\be
\label{constraint}
\sum_{j=1}^{n^\prime} \Phi_{\alpha j} =0\ . 
\ee
In summary, for this choice of basis, the $Y_\nu$ matrix elements
\be\label{standardynu}
(Y_\nu)_{\alpha j} = u_{\alpha j}\, e^{i\Phi_{\alpha j}}\ ,
\ee
depend on only $n(n^\prime-1)$ physical high-energy phases 
$\Phi_{\alpha j}$, $\alpha \ne j$, since the phases $\Phi_{\alpha j}$, 
$\alpha =j$, are not independent and are determined in terms of these phases
using Eq.~(\ref{constraint}).
of the $n(n^\prime-1)$ 

The basis with $Y_e$ and $M$ both diagonal and real, and with $Y_\nu$ given by
Eqs.~(\ref{constraint}) and (\ref{standardynu}) will be called the 
standard high-energy basis.  
In this basis, the $\zeta_\alpha$ are chosen to be independent of the 
$n(n^\prime-1)$ physical phases, which is the simplest choice.  
Alternative, 
more complicated, bases are possible.  One class
of these bases is presented in Appendix 1. Another class includes the physical 
low-energy basis, for which the $\zeta_\alpha$ have a dependence on the physical 
high-energy phases.  This low-energy basis will be 
described later in the paper in subsection $3.3$.

In general, a given high-energy theory is described initially by arbitrary
matrices $M$, $Y_\nu$ and $Y_e$. When rewriting the theory in the standard 
high-energy basis, all physical phases, including those stemming from the 
Majorana character of the heavy neutrinos contained initially in $M$, are 
transferred to the matrix $Y_\nu$ by the diagonalization of $M$ and $Y_e$.

Leptogenesis occurs at high energy in the seesaw model 
due to the presence of $CP$-violation in
the decays of the heavy Majorana neutrinos $N_i$ to light leptons and Higgs
bosons~\cite{fy,luty}.
The $CP$ asymmetry produced in the decay of $N_i$
is given by
\be
\label{leptoasym}
\epsilon_i = {1 \over {\pi \left( Y_\nu^\dagger Y_\nu \right)_{ii}}}
 \sum_{j \ne i}^{n^\prime}
\left\{ \mrm{Im}\left[ \left( Y_\nu^\dagger Y_\nu \right)_{ij}
\right]^2 \,f\left(\frac{|M_j|^2}{|M_i|^2}\right) \right\} \ .
\ee
where $f$ is the loop function \cite{fy,luty,loop}
\be
f(x) = \sqrt{x}\left[\frac{2-x}{1-x} - (1+x) \ln\left(\frac{1+x}{x}\right)\right]\ .
\ee

Notice that the $CP$-asymmetries are invariant under $V_\ell$ transformations,
since both $M$ and the matrix elements 
\be
\label{ylepto}
\left( Y_\nu^\dagger Y_\nu \right)_{ij} = \sum_{\alpha=1}^{n} u_{{\alpha} i}\, u_{{\alpha} j} \,
e^{i (\Phi_{\alpha j}- \Phi_{ \alpha i})} 
\ee
are invariant under these transformations.  Thus, the lepto-asymmetries are
independent of the low-energy basis.

For non-degenerate heavy Majorana neutrinos, the
lepton asymmetry is generated by decay of the lightest heavy Majorana
neutrino $N_1$, and is proportional to the $CP$-asymmetry $\epsilon_1$.  
In terms of the physical high-energy parameters of the standard basis, 
\be
\epsilon_1 ={1 \over {\pi \left( \sum_\alpha u_{{\alpha}1}^2 \right)}}
\sum_{j \ne 1}^{n^\prime} \, \left\{
\mrm{Im}\left[ \left( \sum_{\alpha=1}^n u_{{\alpha}1}\, u_{{\alpha}j}\,e^{i (\Phi_{\alpha j}-\Phi_{\alpha 1})}
 \right)^2\right]
\ f\left(\frac{|M_j|^2}{|M_1|^2}\right)\right\}\, .
\ee

If leptogenesis in the seesaw model produces the
matter-antimatter asymmetry of the universe, then the parameter $\epsilon_1$
is constrained by the experimental value of the baryon-to-photon ratio 
\cite{WMAP},
\be
\eta_B = (6.1 \ {}^{+0.3}_{-0.2})\times 10^{-10} ~,
\label{etaB}
\ee
which is related to the lepto-asymmetry by \cite{BDP}
\be
\eta_B \simeq 0.01  \,\epsilon_1 \cdot \kappa~,
\ee
where the factor $\kappa$ describes the washout of the 
produced lepton asymmetry due to lepton number-violating processes. For 
$M_1 \ll M_2, \, M_3, \cdots, M_{n^\prime}$, the value of $\eta_B$ required 
by experiment typically is obtained when $\epsilon_1 \simeq (10^{-6}-10^{-7})$ 
and $\kappa \simeq (10^{-3}-10^{-2})$.  It is possible to obtain a lower bound 
on $M_1$ from the experimental value of $\eta_B$ and neutrino 
oscillation data \cite{BDP},
\be
M_1 > 4\times 10^8 \ \gev ~.
\ee
This bound indicates a seesaw scale much higher
than the scale of electroweak symmetry breaking, and gives an
estimate for the magnitude of the $1/M$ suppression factor of the $d=6$ operator 
relative to the $d=5$ operator at low energies, provided 
seesaw leptogenesis generates the matter-antimatter
asymmetry of the universe.

\section{Low-energy Effective Lagrangian}

After spontaneous symmetry breaking of the electroweak gauge symmetry, the
effective Lagrangian for leptons is given in the flavor eigenstate basis by
\bea
\label{Leff}
\left({\cal L}_{eff}\right)_{\rm leptons}
&=&i\,\overline e_{L \alpha}\,\D \,e_{L \alpha}+
i\,\overline e_{R \alpha}\,\D \,e_{R \alpha}
- m_{e_\alpha} \left(\overline{e_L}_\alpha\,{e_R}_\alpha+
\overline{e_R}_\alpha\,{e_L}_\alpha \right)+ {\cal L}_{\rm neutrino}\nn\\ 
&&+ {\cal L}_{\rm W,Z} + {\cal L}_{\rm Higgs}\,, 
\eea
where 
the covariant derivative of the charged leptons contains the electromagnetic
gauge field; the charged lepton mass matrix 
$(m_e)= {v \over\sqrt{2}} (Y_e)$ is diagonal with real eigenvalues
$m_{e_\alpha}$ in the flavor basis;
${\cal L}_{\rm neutrino}$ (defined below) contains the kinetic energy and 
Majorana mass terms of the left-handed weakly-interacting neutrinos;
\be
{\cal L}_{\rm W,Z}= \frac{g}{\sqrt{2}}
\left(J_\mu^{-\, CC}\, W^{+ \mu}
+ J_\mu^{+\, CC}\, W^{- \mu} \right)
+\frac{g}{\cos\theta_W}J_\mu^{NC}\, Z^\mu
\ee
contains the $W$ and $Z$ couplings to 
the weak charged and neutral currents, and
${\cal L}_{\rm Higgs}$ contains the lepton couplings to the
Higgs boson.  

In the flavor basis, the charged and neutral currents are given by
\bea
J_\mu^{-\, CC}\,&=& \overline e_{L \alpha} \gamma_\mu \nu_{L \alpha}\,,\\
J_\mu^{NC}\,&=& {1 \over 2} \overline \nu_{L \alpha} \gamma_\mu
\nu_{L \alpha} + \left({-{1 \over 2} -s^2\theta_W }\right)
\overline e_{L \alpha} \gamma_\mu e_{L \alpha}
+ \left(-s^2\theta_W \right) \overline e_{R \alpha} \gamma_\mu e_{R \alpha}\,.\nn
\eea
The neutrino couplings to the Higgs in ${\cal L}_{\rm Higgs}$
depend on the higher-dimension $d > 4$ operators of the effective theory.

\subsection{Low-energy ${d\le5}$-Effective Lagrangian}

The neutrino Lagrangian for the effective theory including only the $d= 5$ 
operator of Eq.~({\ref{d5}) is given in the flavor basis by
\be
\label{d5_lagr}
{\cal L}_{\rm neutrino}^{d\le5} =i\,\overline\nu_{L \alpha}\,\dv\, \nu_{L \alpha}
-\frac{1}{2}
\overline{{\nu_L}^c}_\alpha\,m_{\alpha\beta}\,{\nu_L}_\beta
-\frac{1}{2}\overline{{\nu}_L}_\alpha\,m^*_{\alpha\beta}\,
{{\nu_L}_\beta}^c~,
\ee
 where the light neutrino Majorana mass matrix is\cite{weinberg}
defined by 
\be
m\equiv -{v^2 \over {2}} \, c^{d=5}\,.
\ee  

The matrix $m$ is diagonalized by the transformation 
\be
\label{diag}
m \rightarrow V^* \, m \, V^\dagger = m_{\rm diag, real} = 
{\rm diag}(m_1, m_2,\cdots,m_n) ~,
\ee
where the $n\times n$ unitary matrix $V$ contains $n(n-1)/2$ real and
$n(n+1)/2$ imaginary parameters, and $m_i$ denote the real positive 
eigenvalues of the light Majorana neutrinos.  
The light neutrino Majorana mass eigenstates are defined by
\be
\nu_i= V_{i \alpha } \,{\nu_\alpha}_L
+ V^*_{i \alpha }\,{\nu_\alpha}^c_L \ ,
\ee
and satisfy $\nu_i^c = \nu_i$.
The leptonic Lagrangian of the $d\le5$ effective theory can be rewritten in 
terms of the light Majorana neutrino mass eigenstates,  
\bea
{\cal L}^{d\le 5}_{\rm neutrino}=
\frac{1}{2}\overline {\nu}_i \, \left( i\dvr - m_i \right)\, \nu_i~.
\eea
The weak currents also can be rewritten as 
\be
J_\mu^{-\, CC}=
\overline e_{L \alpha} \,\gamma_ \mu \,(V^\dagger)_{\alpha i}\,\nu_{i},\qquad
(J_\mu^{NC})_{\rm neutrino}= {1 \over 2} \overline \nu_i \gamma_\mu \nu_i \, ,
\ee
where $V$ now appears in the leptonic charged current.   
It is customary to reabsorb $n$ phases $\omega_\alpha$ of $V$ into the
the charged lepton fields, $e^{i \omega_\alpha} \, {e_\alpha} \to e_\alpha$. 
The charged current then becomes 
\be
J_\mu^{-\, CC}=
\overline e_{L \alpha} \,\gamma_ \mu \,U_{\alpha i}\,\nu_{i},\qquad
\ee
where the lepton mixing matrix 
\be
\label{U}
U \equiv \Omega \,  V^\dagger ~~,~~\Omega={\rm diag}(e^{i\omega_1},....
.,e^{i\omega_n})\,,
\label{Omega}
\ee
is the PMNS matrix \cite{PMNS}, which can be written in the form 
\be
\label{UPMNS}
U =U_{CKM} \,D_{Maj}~,
\ee
where $U_{CKM}$ is the leptonic analogue of the $CKM$ matrix of the 
quark sector with $(n-2)(n-1)/2$ Dirac phases and $D_{Maj}$ is a diagonal 
matrix of $(n-1)$ Majorana
phase differences. For two lepton generations, 
\be U=\left(\begin{array}{cc}
\cos\theta&  \sin\theta\, \\
-\sin\theta\,& \cos\theta \\
\end{array}\right)\left(\begin{array}{cc}
e^{i\phi/2}& 0 \\
0& 1 \\
\end{array}\right)\ 
\label{U_2}
\ee
contains a single Majorana phase difference $\phi$ and no $CKM$ phase.  
For three
lepton generations,
\be
U= U_{CKM}
\left(\begin{array}{ccc}
e^{i\phi_{1}/2}& 0 & 0\\
0& e^{i\phi_{2}/2}& 0\\
0& 0 & 1  \\
\end{array}\right)\ 
\ee
contains two Majorana phase differences $\phi_{1}$ and $\phi_{2}$ in
$D_{Maj}$, and a single $CKM$ phase $\delta$ in 
\be
U_{CKM}=\left(\begin{array}{ccc}
c_{12} c_{13} & s_{12} c_{13} & s_{13}\, e^{-i\delta} \\
-s_{12} c_{23} - c_{12} s_{23} s_{13} \, e^{i\delta}
&c_{12} c_{23} - s_{12} s_{23} s_{13} \, e^{i \delta}
&s_{23} c_{13} \\
s_{12} s_{23} - c_{12} c_{23} s_{13} \, e^{i\delta}
&-c_{12} s_{23} - s_{12} c_{23} s_{13} \, e^{i\delta}
&c_{23} c_{13} \\
\end{array}\right) \ .
\ee

The general formula for the neutrino-neutrino oscillation probability
is
\bea
\label{P_nunu}
P(\nu_\alpha \to \nu_\beta) = \sum_{ij} U_{\alpha i}^* \, U_{\beta i} \,
U_{\alpha j} \, U_{\beta j}^* \, e^{i(\vec{p_i} - \vec{p_j}) \vec x} \,,
\eea
whereas the general formula for the antineutrino-neutrino oscillation 
probability is
\bea
\label{P_anunu}
P(\nu_\alpha \to \bar\nu_\beta) = \sum_{ij} U_{\alpha i}^*\, U_{\beta i}^* \,
U_{\alpha j} \, U_{\beta j} \,
{{m_i m_j} \over p^2}
e^{i(\vec{p_i} - \vec{p_j}) \vec x} \ .
\eea
As is well-known, \eq{P_nunu} does not depend on $D_{Maj}$, while
\eq{P_anunu} does.

\subsection{Low-energy ${d\le6}$-Effective Lagrangian}

The neutrino Lagrangian for the effective theory including only
$d \le 6$ operators is given by
\be
\label{d6_lagr}
{\cal L}_{\rm neutrino}^{d\le6}= i\,\overline\nu_{L \alpha}\,\dv \,
\left(\delta_{\alpha \beta}+ \lambda_{\alpha \beta} \right) \, \nu_{L \beta}
-\frac{1}{2}
\overline{{\nu_L}^c}_\alpha\,m_{\alpha\beta}\,{\nu_L}_\beta
-\frac{1}{2}\overline{{\nu}_L}_\alpha\,m^*_{\alpha\beta}\,
{{\nu_L}_\beta}^c~,
\ee
where $m$ is defined as before, and 
\be
\lambda \equiv {v^2 \over {2}} \, c^{d=6}
\ee
is the contribution of the
$d=6$ operator of \eq{d6} to the left-handed neutrino kinetic energy, which is flavor 
non-diagonal.  We can rotate to the basis in which both the neutrino kinetic energy and 
mass matrices are diagonalized, and the neutrino field is rescaled so that
the neutrino kinetic energy is normalized. In this basis, the light neutrino
Majorana mass eigenstates are:
\be
\nu_i= V_{i \alpha }^{\rm{eff}} \,{\nu_\alpha}_L
+ V^{\rm{eff}\,*}_{i \alpha }\,{\nu_\alpha}^c_L~,
\ee
where
\bea
V^{\rm{eff}}\equiv V\,\left({\mathbb I}-\frac{\lambda}{2}\right)
\eea
depends on the unitary matrix $V$
as well as on the factor 
$\left(\delta_{\alpha\beta}-\frac{\lambda_{\alpha\beta}}{2}\right)$
which rescales the neutrino field in the flavor basis. $V^{\rm eff}$ is 
the matrix which diagonalizes the light neutrino Majorana mass 
matrix\footnote{Notice that $m$ itself is 
not affected by the rescaling, to this order in the $1/M$ expansion.},
\be
\left(V^{\rm eff}\right)^* \, m \, \left(V^{\rm eff}\right)^\dagger = m_{\rm diag, real}~.
\ee
Unlike $V$, 
$V^{\rm{eff}}$ is not a unitary matrix because of the field rescaling.
The light Majorana mass eigenstates continue to satisfy $\nu_i^c = \nu_i$.

The leptonic Lagrangian of the $d\le 6$ effective theory can be rewritten in
terms of the light Majorana neutrino mass eigenstates.  The free
neutrino Lagrangian becomes
\bea
{\cal L}_{\rm neutrino}^{d\le 6}=
\frac{1}{2}\overline {\nu_i} \left( i\dvr - m_i \right) {\nu_i}~.
\eea
The weak currents, written in terms of the Majorana mass eigenstates, are
\bea
\label{JCC_d6}
J_\mu^{-\,CC}  
&\equiv& \overline {e_L}_\alpha \, \gamma_\mu \,
({V^{\rm{eff}}}^\dagger)_{\alpha i} \, \nu_i,\\
\label{JNC_d6}
J_\mu^{NC}
&\equiv& {1 \over 2} \overline \nu_i \,\gamma_\mu
({V^{\rm{eff}}})_{i \alpha}\,({V^{\rm{eff}}}^\dagger)_{\alpha j}
\,\nu_j ,
\eea
where ${V^{\rm{eff}}}^\dagger$ appears in the charged current and a factor
${V^{\rm{eff}}_{i \alpha}}\,{V^{\rm{eff}}_{\alpha j}}^\dagger \ne
\delta_{ij}$ appears in the neutral current since $V^{\rm{eff}}$ is not
unitary.

By absorbing the $n$ phases $\omega_\alpha$ into the charged lepton fields as before, the physical
weak currents are described by Eqs.~(\ref{JCC_d6}) and~(\ref{JNC_d6}) 
with the substitution ${V^{\rm{eff}}}^\dagger \to U^{\rm{eff}}$, where 
\be
\label{Ueff}
U^{\rm eff}= \left({\mathbb I}- \Omega \,
\frac{\lambda}{2} \, \Omega^\dagger  \right)
\,U\,.
\ee

The oscillation formulae are now given by Eqs.~(\ref{P_nunu}) and~(\ref{P_anunu}) 
with the substitution $U\to U^{\rm {eff}}$.

\subsection{Global symmetries}

The kinetic energy terms of the Standard Model theory are invariant under
the chiral symmetry transformations
\be
\nu_L \to V_\ell \nu_L,\, \qquad e_{L} \to V_\ell \, e_L, 
\qquad e_R \to V_e \, e_R,
\ee
where $V_\ell$ and $V_e$ are $n \times n$ unitary matrices for $n$ generations
of light fermions.  The chiral symmetry group $G = U(n)_\ell \times U(n)_e$ is 
completely broken by the charged lepton mass matrix $m_e$, the light Majorana 
neutrino mass matrix $m$, and the light neutrino kinetic energy matrix 
$({\mathbb I}+\lambda)$.  
The mass and kinetic energy terms would respect the chiral symmetry if the 
matrices $m_e$, $m$ and $\lambda$ transformed as 
\bea
\label{LE_chiral}
m_e &\to& V_\ell \,m_e\, V_e^\dagger\,,\nn\\
m &\rightarrow& V_\ell^* \, m \, V_\ell^\dagger ~,\ \\
\lambda &\rightarrow& V_\ell \, \lambda \, V_\ell^\dagger ~,\nn
\eea
under the chiral symmetry.  As shown in Ref.~\cite{bgj}, the number of physical parameters of the low-energy
effective theory including the $d=5$ and $d=6$ operators 
is equal to the number of parameters in the matrices
$m_e$, $m$ and $\lambda$ 
minus the number of parameters in $V_\ell$ and $V_e$.

Any high-energy basis with $Y_e$ diagonal and real produces a diagonal and real
charged lepton mass matrix $m_e$ in the spontaneously broken low-energy
theory,
\be
\left( m_e \right)_{\rm diag, real} = {\rm diag}\left( m_{e}, m_{\mu}, \cdots,
m_{e_n} \right) \ .
\ee
This charged lepton mass matrix is left invariant under
$V_\ell = V_e$ rephasings Eq.~(\ref{V_ell}) which depend on $n$ imaginary parameters, $\zeta_\alpha$, $\alpha = 1, \cdots, n$.  

In a general basis, 
the complex symmetric $n \times n$
matrix $m$ contains $n(n+1)/2$ real and $n(n+1)/2$ imaginary parameters defined
by its matrix elements
\be
m_{\alpha \beta} = |m_{\alpha \beta}|\, e^{i \gamma_{\alpha \beta}} \ ,
\ee 
whereas the
Hermitian $n \times n$ matrix $\lambda$ contains $n(n+1)/2$ real and
$n(n-1)/2$ imaginary parameters defined by its matrix elements
\be
\lambda_{\alpha \beta} = |\lambda_{\alpha \beta}|\, 
e^{i \sigma_{\alpha \beta}} \ .
\ee
Under $\zeta_\alpha$ rephasings, the phases
\bea
\gamma_{\alpha \beta} &\to& \gamma_{\alpha \beta}- 
\left(\zeta_\alpha +\zeta_\beta \right) \, ,\\
\sigma_{\alpha \beta} &\to& \sigma_{\alpha \beta}+ 
\left(\zeta_\alpha -\zeta_\beta \right) \, .
\eea

The low-energy $d\le6$ effective theory contains
$n(n-1)$ physical, basis-independent, phases:
\bea
\overline\gamma_{\alpha \beta} &\equiv& \gamma_{\alpha \beta}- 
\frac{\left(\gamma_\alpha +\gamma_\beta \right)}{2} \, ,\\
\overline\sigma_{\alpha \beta} &\equiv& \sigma_{\alpha \beta}+ 
\frac{\left(\gamma_\alpha -\gamma_\beta \right)}{2} \, , 
\eea
where $\gamma_\alpha \equiv \gamma_{\alpha \alpha}$ and
$\gamma_\beta \equiv \gamma_{\beta \beta}$ are the diagonal phases
of the light neutrino Majorana mass matrix $m$ in the flavor basis.
The physical phases $\overline\gamma_{\alpha \beta}$ and $\overline\sigma_{\alpha \beta}$ are invariant under $\zeta_\alpha$ rephasings.

A physical low-energy basis expressed only in terms of the physical 
low-energy phases $\bar \gamma_{\alpha \beta}$ and $\bar \sigma_{\alpha \beta}$ is then defined by the choice of basis 
\be
\zeta_\alpha = \gamma_{\alpha}/2\,.  
\label{lezeta}
\ee
In the physical low-energy basis, the light neutrino mass matrix $m$ depends on 
the $n(n-1)/2$ phases $\overline\gamma_{\alpha \beta}$, $\alpha \ne \beta$, and 
the $\lambda$ matrix depends on the $n(n-1)/2$ phases $\overline\sigma_{\alpha \beta}$,
$\alpha \ne \beta$.
All in all, the basis contains a total of $n(n+2)$ real and $n(n-1)$
imaginary physical parameters in the matrices $m_e$, $m$ and $\lambda$.
The number of physical low-energy parameters is equal to the number of 
physical parameters of the high-energy seesaw model for $n=n^\prime$~\cite{bgj}.

It is worth emphasizing that the physical low-energy basis is not identical
to the standard high-energy basis, $\zeta_\alpha=-\Psi_\alpha$ . 
The former, $\zeta_\alpha = \gamma_\alpha/2$, is related to the 
latter by a relative $V_\ell=V_e$ transformation 
with $\zeta_\alpha = \Psi_\alpha + \gamma_\alpha/2$.  It is worth noting
that these last phases are invariant under $V_\ell=V_e$ rephasings, 
as expected for a rephasing relating two physical
bases: the standard high-energy and the physical low-energy bases.  
 
\section{Connection between High- and Low-Energy Parameters}
\label{connection}

The matrices $M$ and $Y_\nu$ of the high-energy seesaw model
can be determined in terms of the $d=5$ and $d=6$ coefficient matrices
$m$ and $\lambda$ of the broken low-energy effective theory if $n^\prime \le n$.
In this section, we derive explicit formulae for this connection.  We first
derive formulae which produce $M$ and $Y_\nu$ in unstandard form.  
The formulae for $M$ and $Y_\nu$ in the standard high-energy basis are
related to them by a $V_N$ transformation.    

The definitions
\bea
\label{c_d5}
m &\equiv& -{v^2 \over 2} c^{d=5} = -{v^2 \over 2} \left[  
Y_\nu^* \,(M^*)^{-1} \,Y_\nu^\dagger \right]\,,
\\
\label{c_d6}
\lambda &\equiv& {v^2 \over 2}\, c^{d=6} 
= {v^2 \over 2}\, \left[
(Y_\nu \, M^{-1}) (Y_\nu \, M^{-1} )^\dagger \right] \,,
\eea
express the low-energy matrices $m$ and $\lambda$ in terms of the high-energy
matrices $Y_\nu$ and $M$, in a basis-independent way. 

For further use, we define the quantity $\chi$ through 
\be\label{epsxi}
\lambda =\lambda^\dagger= \chi \chi^\dagger.
\ee
One possible solution for $\chi$ is
\bea
\label{xi}
\chi &\equiv&  -{ v \over \sqrt{2}} \,Y_\nu M^{-1} \,.
\eea 
Consistency with the chiral transformations given in Eqs.~(\ref{chiral}) 
and~(\ref{LE_chiral}) implies that 
\bea
\chi  &\to& V_\ell \, \chi \, V_N^T\,.
\eea
under the chiral symmetry.

Using Eqs.~(\ref{c_d5}) and~(\ref{xi}), it is possible to solve for
the high-energy matrices $Y_\nu$ and $M$ in terms of the
low-energy matrices $m$ and $\chi$,
\bea
\label{Ynu_m_epsilon}
Y_\nu &=& {\sqrt{2} \over v} \ m^\dagger \, (\chi^{-1})^T \, ,\\
\label{M_m_epsilon}
M &=& - \,(\chi^{-1})\, {m^\dagger} \, (\chi^{-1})^T\,. 
\eea
This model-independent determination of the seesaw parameters $Y_\nu$ and $M$ 
requires the knowledge of both $m$ and $\chi$, and thus it is only possible when 
both the $d=5$ and $d=6$ coefficients are measured at low energies.  
\emph{A priori}, one can deduce all of the physical high-energy parameters 
from hypothetical measurements of the low-energy parameters.  The caveat refers 
to the practical difficulty of measuring the $d=6$ coefficients, for natural values 
of the seesaw scale.  As we discuss in detail momentarily, the matrices $Y_\nu$ and
$M$ given by Eqs.~(\ref{Ynu_m_epsilon}) and~(\ref{M_m_epsilon}) do not correspond
to the standard high-energy basis.
 
The lepto-asymmetry parameter which is relevant for leptogenesis depends on
the high-energy matrices $M$ and $Y_\nu$.  In a general basis, it depends on 
both the phases of the heavy Majorana neutrino mass matrix $M$, 
Eq.~(\ref{M_m_epsilon}), and the phases of the matrix elements of
\bea
\label{lepto_comb}
Y_\nu^\dagger Y_\nu &=&{2 \over v^2}\,  \,(\chi^{-1})^{*} m m^\dagger
(\chi^{-1})^{T}\,  .
\eea
Notice that this expression is invariant under $V_\ell$ transformations, while 
it depends on the high-energy basis chosen for the heavy Majorana neutrinos,
transforming under $V_N$ transformations as
\bea
\label{lepto_trans}
Y_\nu^\dagger Y_\nu \to V_N\,Y_\nu^\dagger \,Y_\nu  \,V_N^\dagger\,.
\eea

There is an important subtlety to the above derivation of the high-energy matrices.
The explicit expressions Eqs.~(\ref{Ynu_m_epsilon}) 
and~(\ref{M_m_epsilon}) for $M$ and $Y_\nu$ 
depend on the low-energy basis chosen for $m$ and $\lambda$.  
For instance, when the matrices $m$ and $\lambda$ are 
written in the physical low-energy basis, the heavy neutrino mass matrix $M$ obtained 
using Eq.~(\ref{M_m_epsilon}) is neither diagonal nor real, as it is, by definition, 
in the standard high-energy basis. A further $V_N$ transformation is required to obtain the standard high-energy basis with $M$ diagonal and real. 

The point is that the definition of $\chi$ given in Eq.~(\ref{xi})
is not unique.  It is clear that multiplying the right-hand side of 
Eq.~(\ref{xi}) by any unitary matrix,
\be
\chi \rightarrow \chi \,{V_N}^T ,
\label{ay}
\ee
 results in an equally valid solution for $\chi$,
since it produces the same $\lambda$ matrix as Eq.~(\ref{xi}).  
The unitary matrix is denoted by $V_N$ to maintain 
consistency with the chiral transformations given in Eqs.~(\ref{chiral}) 
and~(\ref{LE_chiral}). Different $V_N$ transformations correspond to 
different choices of the high 
energy basis, as  $Y_\nu \rightarrow Y_\nu {V_N}^\dagger$ and 
$M \rightarrow {V_N}^*M {V_N}^\dagger$.  The specific $V_N$ which makes 
$M$ diagonal and real yields the standard high-energy basis.

Let us consider more explicitly the low-energy basis.
The $m$ matrix is diagonalized by the unitary matrix $V$ as defined
in Eq.~(\ref{diag}).  Let us define
the unitary matrix $W$ which diagonalizes the $\lambda$ matrix by
\be
\label{lambdadr}
\lambda_{\rm diag, \, real} \equiv W^\dagger \,\lambda W\, .
\ee  
From Eq.~(\ref{epsxi}), the general solution for $\chi$ is given by
\bea
\label{xi2}
\chi &=& W \, \sqrt{\lambda_{\rm diag, \, real}}\, V_N^T\,.
\eea
$Y_\nu$ and $M$ and the matrix $Y_\nu^\dagger Y_\nu$ relevant for leptogenesis 
now can be expressed in all generality as\footnote{ The results in this section can be 
easily rewritten in terms of the usual PMNS matrix, 
with the help of \eq{U}. This is done for the two generation case in section 5.}
\bea
\label{Ynu_L}
Y_\nu &=& {\sqrt{2} \over v} \ V^\dagger \, m_{\rm diag,\, real} \,L,\\
\label{M_L}
M  &=& - L^T \, m_{\rm diag, real} \,L\, ,\\
\label{lepto_L}
Y_\nu^\dagger Y_\nu &=&{2 \over v^2}\, L^\dagger \, m^2_{\rm diag,real} \, L~,
\eea 
where the matrix $L$ is given by
\bea
\label{L}
L\equiv   \,(\chi^{-1}\,V^\dagger)^T\,=
\,
 V^* \, W^* \,
(\sqrt{\lambda_{\rm diag, \, real}})^{-1}\,  \, V_N^\dagger\,.
\eea
The standard high-energy basis is obtained if
$V_N$ is chosen to be the unitary matrix which 
diagonalizes $M$ in Eq.~(\ref{M_L}).
Thus, $V_N$ is
determined {\it a priori} from low-energy measurements, 
\bea
\label{diag_M}
M_{\rm diag,\,real}&\equiv& -V_N^*\,\left((\sqrt{\lambda_{\rm diag, \, real}})^{-1} \, 
W^\dagger \,V^\dagger \,m_{\rm diag,\,real}\,V^* 
W^*\, (\sqrt{\lambda_{\rm diag, \, real}})^{-1} \right)\,V_N^\dagger\,.
\eea

It is easy to show a number of interesting results using the solutions
Eqs.~(\ref{Ynu_m_epsilon}) and~(\ref{M_m_epsilon}) for $Y_\nu$ and $M$.  
For example, if there is no $CP$-violation at low energies, then there 
is a low-energy basis in which $m$ and $\lambda$ are both real. 
From the definitions in Eqs.~(\ref{diag}) 
and (\ref{lambdadr}), it follows that $V$ and $W$ are real, as are
$V_N$ in \eq{diag_M} and therefore $\chi$ in Eq.~(\ref{xi2}).
Thus, it follows from Eqs.~(\ref{Ynu_m_epsilon}) and~(\ref{M_m_epsilon})
that $Y_\nu$ and $M$ are real, and there is no leptogenesis at high energies.

Eqs.~(\ref{diag_M}) simplifies considerably for a $\lambda$ 
matrix with equal eigenvalues, $\lambda_i= |\lambda|$. 
The heavy neutrino masses then are given by 
\bea
\label{M_m}
M_{\rm diag,real} &=& 
-\frac{1}{|\lambda|}\, V_N^*\, V^\dagger\, m_{\rm diag,real} \,V^*\, 
V_N^\dagger~=-\frac{m_{\rm diag,real}}{|\lambda|}~,
\eea
where $W=1$, $V_N=V^*$, and Eq.~(\ref{L}) reduces to $L={\mathbb I}/{\sqrt{|\lambda|}}$. 
In other words, 
the heavy neutrino masses are proportional to the light neutrino masses 
when $\lambda$ has degenerate eigenvalues.
Eq.~(\ref{M_m}) further implies that the heavy Majorana neutrino masses become 
degenerate, $|M_i|= |M|$, in the limit when both the light neutrinos 
masses and the $\lambda$-eigenvalues 
become degenerate, $m_i = |m|$ and $\lambda_i = |\lambda|$:
\be
|M|=\frac {|m|}{|\lambda|}\,.
\ee

Let us now turn to the analysis of the lepto-asymmetries in the basis in which $M$ is diagonal and real.
From Eq.~(\ref{lepto_L}), it follows that leptogenesis vanishes in the following two 
limits:
\begin{itemize}
\item{Degenerate eigenvalues $\lambda_i = |\lambda|$ of the $d=6$ matrix 
$\lambda$.  In this limit, the matrix $L={\mathbb I}/\sqrt{|\lambda|}$,
and $Y_\nu^\dagger Y_\nu$ in Eq.~(\ref{lepto_L}) is real 
and diagonal,
\bea
\label{lepto_comb_VN}
Y_\nu^\dagger Y_\nu &=& = {2 \over v^2}\,\frac{m_{\rm diag,real}^2}{|\lambda|}
~.
\eea
}

\item{Degenerate light neutrino masses $m_i = |m|$,    
when the heavy neutrinos are strongly hierarchical.  For  $M_1\ll M_2,M_3,
\cdots, M_{n^\prime}$, the lepto-asymmetry 
$\epsilon_1$ given in Eq.~(\ref{leptoasym}) is approximated by \cite{DI}
\bea
\epsilon_1
&\simeq& -\frac{3}{8\pi} \frac{
|M_1|}{[Y_\nu^\dagger Y_\nu]_{11}}
\sum_{j \ne 1}^{n^\prime} 
\frac{{\rm Im} \,\left[\left(Y_\nu^\dagger Y_\nu\right)_{1j}\right]^2}{|M_j|}\nn\\
&=& -\frac{3}{8\pi} \frac{1}{[Y_\nu^\dagger Y_\nu]_{11}}
\sum_{j \ne 1}^{n^\prime} 
{\rm Im} \,\left\{ {{M_1} \over {M_j}}
\left[\left(Y_\nu^\dagger Y_\nu\right)_{1j}\right]^2\right\}\nn\\
&=&\frac{3}{4\pi} \frac{1}{v^2}\,\frac{1}{
[Y_\nu^\dagger Y_\nu]_{11}}\,{\rm Im}
\{M_1 [Y_\nu^\dagger\, m^\dagger\,Y_\nu^*]_{11}\}\,,
\eea
or, in terms of $L$ using
Eqs.~(\ref{Ynu_L}) and~(\ref{M_L}),
\bea
\label{uf}
\epsilon_1 &\simeq& - 
\frac{3}{4\pi} \,\frac{1}{v^2}\,\frac {{\rm Im}\{[L^T\,m_{\rm diag,real}\,L]_{11}\,[L^\dagger\,m_{\rm diag,real}^3\,L^*]_{11}\}}
{[L^\dagger\,m^2_{\rm diag,real}\,L]_{11}}.
\eea
In the limit when all the light neutrinos have the same mass, i.e.
$m_{\rm diag,\,real} = |m| \, {\mathbb I}$, the $CP$-odd lepto-asymmetry vanishes:
\bea
\epsilon_1 &\simeq& -\frac{3}{4\pi} \,
\frac{|m|^2}{v^2} \,\frac{1}{[L^\dagger\,L]_{11}}\,
{\rm Im}\{[L^T\,L]_{11}\,[L^T \,L]^\dagger_{11}\}= 0\,,
\eea
since the diagonal matrix elements of a Hermitian matrix are real. 

The lepton asymmetry in \eq{uf} can be rewritten in terms of the squared mass 
differences of the light neutrino masses \cite{BDP,DI} 
to emphasize that it vanishes in the limit
of degenerate light neutrino masses,
\bea
\epsilon_1&=& -\frac{3}{4\pi} \,\frac{1}{v^2} 
\frac{[L^T \,m_{\rm diag,real}\,L ]_{11}}{[L^\dagger\,m^2_{\rm
diag,real}\,L]_{11}}\,
\sum_j m_j^3 \,{\rm Im}\{[L^*]^2_{j1}\}\  \\
&=&-\frac{3}{4\pi} \,\frac{1}{v^2} 
\frac{[L^T \,m_{\rm diag,real}\,L ]_{11}}{[L^\dagger\,m^2_{\rm
diag,real}\,L]_{11}}\,
\sum_j m_j\,\Delta m_{j1}^2\,{\rm Im}\{[L^*]^2_{j1}\}\,,
\eea
where we have used
\bea
m_1\,{\rm Im}\{ M_1\} = m_1\,\sum_j m_j\,{\rm Im}\{[L^*]^2_{j1}\}= 0\,.
\eea
}
\end{itemize}

\subsection{Connection with the literature}

Many different parameterizations for the Yukawa couplings of the heavy Majorana 
neutrinos can be found in the literature. 
Our treatment is an improvement over earlier ones in that no matrix with undefined matrix
elements is used.  Instead, all high-energy parameters are derived in terms of
matrix elements of the $d=5$ and $d=6$ operator coefficients.  These matrix elements can 
be determined {\it a priori} from 
low-energy measurements; indeed, we have related all $d=5$ and $d=6$ matrix elements
to $CP$-odd and $CP$-even observables.

A very popular parameterization is that in which $Y_\nu$ is expressed in terms of an 
orthogonal matrix $R$ \cite{Casas}, 
\be
\label{R}
Y_\nu = V^\dagger \sqrt{m_{\rm diag, \, real}}\, 
R^\dagger \, \sqrt{|M_{\rm  diag,\, real}|}\,.
\ee
The combination  relevant for leptogenesis then is written as 
\bea
\label{lepto_comb3}
Y_\nu^\dagger Y_\nu &=&  \sqrt{|M_{\rm  diag,\, real}|}\,R \,m_{\rm diag, \, real}
\, R^\dagger \, \sqrt{|M_{\rm  diag,\, real}|}\,.
\eea
$R$ is left arbitrary in the usual treatment which only includes the
effects of the $d=5$  operator.  Within this treatment, all the
$CP$-violating phases are embedded in $R$, and are, therefore, unknown. 
The ${d\le6}$-effective theory instead allows $R$ to be determined in terms 
of the $d=5$ and $d=6$ coefficients.  From a comparison of 
Eqs.~(\ref{Ynu_L}) and~(\ref{R}), we obtain
\be
R ={\sqrt{2} \over v} \left(\sqrt{|M_{\rm  diag, \, real}|}\right)^{-1}
\,L^\dagger\,
\sqrt{m_{\rm diag, \, real}}
\,.
\ee
Notice that $R$ is real whenever $L$ is real.

In a different approach \cite{PPR2}, called the bi-unitary parameterization, 
matrices $U_L$ and $U_R$ are defined such that
\bea
Y_\nu = U_L^\dagger\,(Y_\nu)_{\rm diag,real}\,U_R\,.
\eea
Comparison with the chiral transformations in Eq.~(\ref{chiral}) shows that $U_L$ 
corresponds to a particular choice of $V_\ell$, while $U_R$ belongs to the class of 
$V_N$ transformations.  Notice, though, that $U_R$ does not correspond to the $V_N$ 
transformation which diagonalizes $M$ but to the diagonalization of 
$Y_\nu^\dagger Y_\nu$, with $M$, in general, remaining complex and non-diagonal. 
From our Eqs.~(\ref{lepto_L}) and (\ref{L}), the dependence of 
$U_R$ and $U_L$ on the low-energy mixing parameters can be extracted.

A different series of works \cite{triang} takes advantage of the properties of triangular 
matrices,  defining 
\bea
Y_\nu= {\bf U}\,Y_\Delta\,,
\eea
where ${\bf U}$ is unitary and $Y_\Delta$ triangular.
${\bf U}$ now can be identified with yet another particular $V_\ell$ 
transformation.

Finally, let us consider the matrix $Y_\nu Y_\nu^\dagger$.  In supersymmetric versions
of the seesaw mechanism, renormalization effects due to the neutrino Yukawa interaction 
induce flavor-mixing terms in the slepton mass matrices~\cite{SUSY}, resulting
in lepton flavor-violating decays~\cite{Casas,LFV}. The relevant
quantities are the off-diagonal matrix elements of $Y_\nu Y_\nu^\dagger$. 
In the minimal, non-supersymmetric, seesaw model, this combination 
can be written in terms of the low-energy observables as
\bea
\label{SUSY_comb}
Y_\nu Y_\nu^\dagger &=&{2 \over v^2}\,    m^\dagger\,(\chi^{-1})^{T}\, \,(\chi^{-1})^{*}\,m \nn\\
&=&{2 \over v^2}\, V^\dagger\,m_{\rm diag,real}\,L\,L^\dagger 
\, m_{\rm diag,real} \, V~.
\eea
In the limit of degenerate eigenvalues of the $d=6$ coefficient matrix, 
the only mixing parameters which remain in \eq{SUSY_comb} are the usual 
neutrino mixing parameters stemming from the $d=5$ operator,
 \bea
Y_\nu Y_\nu^\dagger &=&
{2 \over v^2}\,\frac{1}{|\lambda|}\, V^\dagger\,m^2_{\rm diag,real} \, \, V~.
\eea 
In fact, $Y_\nu Y_\nu^\dagger$ is a quantity which depends on the low-energy 
basis chosen, as can be seen from its chiral transformation properties
under $V_\ell$ rephasings,
\bea
Y_\nu Y_\nu^\dagger \to V_\ell \,Y_\nu Y_\nu^\dagger \,V_\ell^\dagger\,,
\eea
whereas it is invariant under $V_N$ chiral transformations.
This behavior is opposite to that of the combination relevant for leptogenesis 
$Y_\nu^\dagger Y_\nu$, see \eq{lepto_trans}.

\section{Two generations}

In this section, we explicitly relate the parameters of the 
high-energy seesaw model to the low-energy parameters of the effective theory 
in the case of two generations of heavy neutrinos 
and Standard Model leptons. In this case, the leptonic sectors of 
the high-energy seesaw Lagrangian and the low-energy
$d\le 6$-effective Lagrangian each contain 8 real and 2 imaginary parameters.

\subsection{High-energy Seesaw Lagrangian}

The seesaw model with two lepton generations
is defined by $2 \times 2$ matrices $Y_e$, $M$ and $Y_\nu$.
In the standard high-energy basis, $Y_e$ and $M$ are
the diagonal and real matrices
\be
(Y_e)_{\rm diag,\, real}
= \left(\begin{array}{cc}
y_e &0\\
0& y_\mu
\end{array}\right)~ \equiv {\sqrt{2} \over v} \,
\left(\begin{array}{cc}
m_e &0\\
0& m_\mu
\end{array}\right),
\ee
\be
M_{\rm diag, \, real}= 
\left(\begin{array}{cc}
|M_1|\, & 0\\
0 & |M_2|\, 
\label{M}
\end{array}\right)~,
\ee
whereas the matrix $Y_\nu$ is given by
\be
\label{Ynu}
Y_\nu =
\left(\begin{array}{cc}
u_{e1}\, e^{-i \Phi_{e2}} & u_{e2} \, e^{i \Phi_{e2}} \\
u_{\mu 1}\, e^{i \Phi_\mu} & u_{\mu2} \, e^{-i \Phi_\mu} \\
\end{array}\right)\equiv \left(\begin{array}{cc}
u_{1}\, e^{-i \Phi_{e}} & u_{2} \, e^{i \Phi_{e}} \\
v_{1}\, e^{i \Phi_{\mu}} & v_{2} \, e^{-i \Phi_{\mu}} \\
\end{array}\right).
\ee
For simplicity, the notation defined by the right-hand side of this equation will 
be used hereafter instead of the general notation displayed on the left-hand side 
of the equation. 

The 2 physical phases of the high-energy theory are
$\Phi_e$ and $\Phi_\mu$, whereas
the 8 real parameters are $y_e$, $y_\mu$, $M_1$, $M_2$, $u_1$, $v_1$, $u_2$ and $v_2$.
The charged lepton Yukawa parameters $y_e$ and $y_\mu$ are not free parameters, but are
known in terms of the charged lepton masses and 
the Higgs vacuum expectation value.

Leptogenesis at high energy depends on the $CP$ asymmetries produced in heavy Majorana neutrino decay. The $CP$ asymmetry  produced by decay of $N_1$ is 
\bea
\epsilon_1 &=&  {1 \over {\pi \left( Y_\nu^\dagger 
Y_\nu \right)_{11}}}
\mrm{Im}\left[\left({Y_\nu}^\dagger Y_\nu\right)_{12}
\right]^2\,
f\left(\frac{|M_2|}{|M_1|}\right)\,,
\eea
which is equal to 
\be
\epsilon_1 ={1 \over {\pi \left( u_1^2 + v_1^2 \right)}}
\mrm{Im}\left[ \left( u_1u_2\,e^{2i \Phi_e}
+v_1v_2\,e^{-2i \Phi_\mu} \right)^2\right]
\ f\left(\frac{|M_2|}{|M_1|}\right)\,,
\ee
in the standard high-energy basis. 
 
\subsection{$d=5$ and $d=6$ coefficients}

The $d=5$ and $d=6$ operator coefficients of the low-energy effective Lagrangian, 
$c^{d=5}$ and $c^{d=6}$, are defined in terms of the matrices $M$ and $Y_\nu$ of 
the high-energy seesaw Lagrangian by Eqs.~(\ref{cd5}) and~(\ref{cd6})~\cite{bgj}.
In the standard high-energy basis, they are 
\bea
\label{c5_b1}
c^{d=5}&=& \left(\begin{array}{cc}
\frac{u_1^2 }{M_1}\, e^{i \,2\Phi_e}
 +\frac{u_2^2}{M_2}\,e^{-i\,2\Phi_e}&
\frac{u_1\,v_1}{M_1}\,e^{i\,(\Phi_e-\Phi_\mu)}
+\frac{u_2\,v_2}{M_2}\,e^{-i\,(\Phi_e-\Phi_\mu)} \\
\frac{u_1\,v_1}{M_1}\,e^{i\,(\Phi_e-\Phi_\mu)}
+\frac{u_2\,v_2}{M_2}\,e^{-i\,(\Phi_e-\Phi_\mu)} &
\frac{v_1^2}{M_1}\,e^{-i\,2\Phi_\mu}
+\frac{v_2^2}{M_2}\,e^{i\,2\Phi_\mu}
\end{array}\right) \nn
\eea
and
\bea
\label{c6_b1}
c^{d=6}&=&\left(\begin{array}{cc}
\frac{u_1^2}{M_1^2} + \frac{u_2^2}{M_2^2}&
\frac{u_1 v_1}{M_1^2}\,e^{-i\,(\Phi_e+\Phi_\mu)}
+ \frac{u_2 v_2}{M_2^2}\,e^{i\,(\Phi_e+\Phi_\mu)} \\
\frac{u_1 v_1 }{M_1^2}\, e^{i\,(\Phi_e+\Phi_\mu)}
+\frac{u_2 v_2 }{M_2^2} \,e^{-i\,(\Phi_e+\Phi_\mu)} &
\frac{v_1^2}{M_1^2} + \frac{v_2^2}{M_2^2} 
\end{array}\right) \nn\, .
\eea

The low-energy real and imaginary parameters of 
the $d=5$ and $d=6$ coefficient matrices are defined by
\be
 c^{d=5}\equiv  - \left(\begin{array}{cc}
|c^{d=5}_{ee}|\,e^{i\gamma_e} & |c^{d=5}_{e \mu}|\,e^{i\gamma_{e \mu}} \\
  |c^{d=5}_{e \mu}|\,e^{i\gamma_{e \mu}}    & |c^{d=5}_{\mu\mu}|\,e^{i\gamma_\mu} \\
\end{array}\right)~,
\ee
\be
 c^{d=6} \equiv  \left(\begin{array}{cc}
|c^{d=6}_{ee}| & |c^{d=6}_{e \mu}|\,e^{i\sigma_{e \mu}} \\
  |c^{d=6}_{e \mu}|\,e^{-i\sigma_{e \mu}}    & |c^{d=6}_{\mu\mu}| \\
\end{array}\right)~.
\ee
(An explicit minus sign has been introduced into the definition of the
$d=5$ operator coefficient parameters to simplify later formulae for
the light neutrino Majorana mass matrix.)

The low-energy phases
$\gamma_e, \gamma_\mu, \gamma_{e \mu}$ of the $d=5$ coefficient and
$\sigma_{e \mu}$ of the $d=6$ coefficient can be expressed in terms 
of the parameters of the high-energy standard basis by
\bea
\label{gamsig}
\gamma_e&=&\arctan
\left(\left({
{{u_1^2 \over M_1}- {u_2^2 \over M_2}}\over
{{u_1^2 \over M_1}+ {u_2^2 \over M_2}}
}\right)
\tan{2\Phi_e}\right)+ \pi~,
\nn\\
\gamma_\mu&=& \arctan
\left(\left({{{v_2^2 \over M_2}- {v_1^2 \over M_1}}\over
{{v_2^2 \over M_2}+ {v_1^2 \over M_1}}}\right)
\tan{2\Phi_\mu}\right)+ \pi~,
\\
\gamma_{e \mu}&=&\arctan
\left(\left({
{{{u_2 v_2} \over M_2}- {{u_1 v_1}\over M_1}}\over
{{{u_2 v_2} \over M_2}+ {{u_1 v_1}\over M_1}}}\right)
\tan{(\Phi_e-\Phi_\mu)}\right)+ \pi~,\nn\\
\sigma_{e \mu}&=& \arctan\left(\left(
\frac {{{u_2 v_2} \over M_2^2} - {{u_1 v_1} \over M_1^2}}
{{{u_2 v_2} \over M_2^2} + {{u_1 v_1} \over M_1^2}}\right)
\tan(\Phi_e+\Phi_\mu)\right)\ .~\nn
\eea

There are only 2 independent low-energy phases which are invariant under 
$\zeta_\alpha$ rephasings at low energies:
\be
\label{lowphasegam}
\bar\gamma \equiv  \gamma_{e \mu}
- \frac {\gamma_e+\gamma_\mu}{2} \,, \\
\ee
and
\be
\label{lowphasesig}
\bar\sigma \equiv \sigma_{e \mu} + \frac {\gamma_e-\gamma_\mu}{2}\,.
\ee

The phases $\bar\gamma$ and $\bar\sigma$
are the 2 physical phases of the low-energy $d \le 6$-effective Lagrangian 
which cannot be eliminated by a change of basis. 
In the following, when considering the low-energy effective theory, we will work in the basis 
where all the unphysical phases are removed by performing a rephasing 
\bea
V_e=V_\ell= 
\left(\begin{array}{cc}
e^{i \gamma_{e}/2}& 0 \\
0& e^{i\gamma_{\mu}/2} \\
\end{array}\right)
\eea
relative to the standard high-energy basis, so that only the physical phases
$\bar\gamma$ and $\bar\sigma$ appear.
In this physical low-energy basis, the $d=5$ and $d=6$ coefficients are given by 
\be
\label{d5_EB}
 c^{d=5}\equiv - \left(\begin{array}{cc}
|c^{d=5}_{ee}| & |c^{d=5}_{e \mu}|\,e^{i\bar\gamma} \\
  |c^{d=5}_{e \mu}|\,e^{i\bar\gamma}    & |c^{d=5}_{\mu\mu}| \\
\end{array}\right)~,
\ee
\be
\label{d6_EB}
 c^{d=6} \equiv  \left(\begin{array}{cc}
|c^{d=6}_{ee}| & |c^{d=6}_{e \mu}|\,e^{i\bar\sigma} \\
  |c^{d=6}_{e \mu}|\,e^{-i\bar\sigma}    & |c^{d=6}_{\mu\mu}| \\
\end{array}\right)~.
\ee

We will see that this basis is particularly useful for relating high- and 
low-energy observables. In particular, explicit formulae that express the 
2 $CP$-odd phases \emph{a priori} measurable in the experiments in terms of 
$\bar\gamma$ and $\bar\sigma$ are given later. Through 
Eqs.~(\ref{gamsig})-(\ref{lowphasesig}), 
the high-energy phases $\Phi_e$ and $\Phi_{\mu}$ can
be directly related to the observable low-energy $CP$-odd phases.

\subsection{Low-energy Effective Lagrangian}

The parameters of the low-energy effective theory including the $d=5$ and
$d=6$ operators are given by the two charged lepton masses $m_e$ and $m_\mu$,
the two light neutrino Majorana masses $m_1$ and $m_2$, the mixing
angle $\theta$, the phase 
of the lepton mixing matrix, the 3 magnitudes
$\lambda_{ee}$, $\lambda_{e \mu}$ and $\lambda_{\mu\mu}$,
and the phase 
of the $\lambda$ matrix.  These parameters are defined
in this subsection.   

The light Majorana neutrino masses and lepton mixing matrix are derived first
in the effective theory including only the $d=5$ operator, 
and then in the effective theory including both the $d=5$ and $d=6$ operators.

\subsubsection{Low-energy ${d\le5}$-Effective Lagrangian}

Consider the neutrino Lagrangian in \eq{d5_lagr}.  In the physical low-energy
basis, the complex symmetric $2\times2$
mass matrix of the light neutrinos depends on 3 real and 1 imaginary parameters:
\bea
m \equiv
-\frac{v^2}{2}\,c^{d=5} &=&
\left(\begin{array}{cc}
m_{ee}& m_{e \mu}\,e^{i\bar\gamma} \\
m_{e \mu}\,e^{i\bar\gamma}& m_{\mu\mu}\\
\end{array}\right)\,~,
\label{m}
\eea
where the $m_{\alpha \beta}$ are real and positive. 
The light neutrino Majorana mass matrix is diagonalized by the transformation
\be\label{22diag}
m \rightarrow V^* \, m \, V^\dagger = m_{\rm diag, real} = \left(\begin{array}{cc}
m_1 & 0 \\
0& m_2 \\
\end{array}\right)~,
\ee
where the $2\times2$ unitary matrix $V$ contains 1 real and 3 imaginary
parameters which are parameterized  by a real mixing angle and 3 phases
\bea
\label{V}
V^\dagger &=&\left(\begin{array}{cc}
e^{-i\hat\theta_{e}/2}& 0 \\
0& e^{-i\hat\theta_{\mu}/2} \\
\end{array}\right)
\,
\left(\begin{array}{cc}
\cos\theta& \sin\theta\,e^{-i\phi/2}\\
-\sin\theta\,e^{i\phi/2}& \cos\theta \\
\end{array}\right)~.
\eea 
The mixing angle $\theta$ is taken to be in the interval $[0,\pi]$.

A discussion of the diagonalization of the low-energy $2\times2$ Majorana 
mixing matrix can be found in Ref.~\cite{Barroso}. There, the unitary matrix 
is given by \footnote{
The phases $\bar\gamma$ and $\rho$, and the mixing angle $\theta$
 defined in this work 
correspond to the phases $\beta$ and $-\alpha$, and the angle $-\theta$
 in Ref.~\cite{Barroso}.}
\bea
\label{Vrho}
V^\dagger &=& 
\left(\begin{array}{cc}
\cos\theta& \sin\theta\,e^{-i\rho  }\\
-\sin\theta\,e^{i\rho  }& \cos\theta \\
\end{array}\right)
\,
\left(\begin{array}{cc}
e^{-i\hat\theta_{e}/2}& 0 \\
0& e^{-i\hat\theta_{\mu}/2} \\
\end{array}\right)~,
\eea
with
\bea
\label{rho}
\rho\equiv\frac{\phi+\hat\theta_e-\hat\theta_\mu}{2}\,.
\eea

The 1 real and 3 imaginary parameters in $V$ are determined by the
diagonalization of the light neutrino Majorana mass matrix, \eq{22diag}.
Under a $V_\ell=V_e$ rephasing, the three phases $ \phi$ (or $\rho$), 
$\hat\theta_e$ and $\hat\theta_\mu$ are invariant.

The $\hat\theta_\alpha$ phases are not directly measurable when considering only the 
$d\le5$ effective Lagrangian, since they can be removed from the 
Lagrangian by performing a rephasing on the charged lepton fields, Eq.~(\ref{Omega}),
\bea
\label{reph_charg_lep}
\Omega \,
\left(\begin{array}{c}
e\\
\mu\\
\end{array}\right)_{L,R}
=\left(\begin{array}{cc}
e^{i\omega_{e} }&0 \\
0& e^{i\omega_{\mu}}\\
\end{array}\right)\,
\left(\begin{array}{c}
e\\
\mu\\
\end{array}\right)_{L,R}
\to
\left(\begin{array}{c}
e\\
\mu\\
\end{array}\right)_{L,R}
\,~,
\eea
which leaves all terms in the leptonic Lagrangian invariant, with the exception
of the weak charged current.
The particular choice 
\bea
\label{zeta_i}
\omega_{e } &=& \left(\hat\theta_e + \phi\right)/{2},\qquad
\omega_{\mu} = {\hat\theta_{\mu}/2}~,
\eea
yields the weak charged current 
\be
J_\mu^{-\, CC}= \overline
e_{L \alpha} \,\gamma_ \mu \, U_{\alpha i}\,\nu_{i}
\ee
in terms of the usual leptonic mixing matrix 
\be U
=\left(\begin{array}{cc}
\cos\theta&  \sin\theta\, \\
-\sin\theta\,& \cos\theta \\
\end{array}\right)\left(\begin{array}{cc}
e^{i\phi/2}& 0 \\
0& 1 \\
\end{array}\right)\ .
\label{U_22}
\ee

It will not be possible to eliminate the two $\hat\theta_\alpha$ phases when 
the $d=6$ operator is included in the low-energy effective Lagrangian.  
Indeed, we show below that the $\hat\theta_\alpha$ phases are essential for 
relating the physical low-energy phases $\bar\gamma$ and $\bar\sigma$ with the $CP$-odd phases 
measurable in experiments.  We will see in the next subsections that the  
$\hat\theta_\alpha$ phases are functions of the mass eigenvalues $m_1$ and 
$m_2$, the mixing angle $\theta$ and the phase $\phi$ (or $\rho$), which is as 
expected since the original matrix $m$ only contained one physical phase. 

Note that it is not
possible to remove the phase $\phi$ (or, equivalently, $\rho$ in the alternative
parameterization of $V$) from the effective theory
by rephasings on the $\nu_i$,
since these rephasings destroy the relations $\nu_i^c = \nu_i$, 
introducing additional
phase factors into these 
relations,
i.e. $\nu_1^c = e^{-i\phi} \, \nu_1$.
These phase
factors contribute to physical observables, yielding results equivalent
to keeping $\phi$ in the lepton mixing matrix.  Throughout we stick to the
convention that $\nu_i^c = \nu_i$, so that rephasings of the 
light Majorana neutrinos are forbidden.

\subsubsection{Mass vs. Flavor Eigenstate Basis at Low Energy} 

Here, we give a brief summary of the relationship of
the mass matrix parameters in the flavor eigenstate and mass eigenstate
bases.  From \eq{diag},
\bea
m_{ee}\,&=& \left[ m_1\,\cos^2\theta\,+
m_2\,\sin^2\theta\,e^{i\phi}\right]\,e^{i\hat\theta_e}\,,
\label{m_ee}\\
m_{\mu\mu}\,&=&\left[m_1\,\sin^2\theta\,e^{-i\phi}+
m_2\,\cos^2\theta\right]\,e^{i\hat\theta_\mu}\,,
\label{m_mumu}\\
m_{e \mu}\,e^{i\,\bar\gamma}&=&(m_2\,e^{i\phi/2}-m_1\,e^{-i\phi/2})
\,\cos\theta\,\sin\theta\,e^{i\,\frac{\hat\theta_e+\hat\theta_\mu}{2}} .
\label{m_emu}
\eea

From the vanishing of the imaginary parts of the r.h.s. 
of Eqs.~(\ref{m_ee}) and~(\ref{m_mumu}), we deduce the following relations
between the phases $\hat\theta_\alpha$ and the physical observables
\bea
\cot \hat\theta_e&=&- \cot \phi - \frac{m_1}{m_2}\,
\frac{\cot^2\theta}{\sin\phi}\,,\\
\cot \hat\theta_\mu&=& \cot \phi + \frac{m_2}{m_1}\,
\frac{\cot^2\theta}{\sin\phi}\,.
\eea
Thus, the phases $\hat\theta_\alpha$ are functions of the four independent 
parameters $m_1$, $m_2$, the mixing angle $\theta$ and the phase $\phi$. 
Notice that the phase $\rho$ of the parameterization in 
Eqs.~(\ref{Vrho}) and~(\ref{rho}), is also a function of these parameters, 
and in consequence, it can be determined only after all of the parameters
have been measured in low-energy experiments.

The real and imaginary parts of \eq{m_emu} yield
\bea
m_{e \mu} \cos\bar\gamma &=& {1 \over 2}\sin2\theta \left[ (m_2- m_1 )
\cos{\phi \over 2} \cos\frac{\hat\theta_e + \hat\theta_\mu}{2} -
(m_1 + m_2) \sin{\phi \over 2}\sin\frac{\hat\theta_e + \hat\theta_\mu}{2}\right],
\nn\\
m_{e \mu} \sin\bar\gamma &=& {1 \over 2}\sin2\theta \left[ (m_2 - m_1 )
\cos{\phi \over 2} \sin\frac{\hat\theta_e + \hat\theta_\mu}{2} +
(m_1 + m_2) \sin{\phi \over 2} \cos\frac{\hat\theta_e 
+ \hat\theta_\mu}{2} \right]\,, \nn\\
\eea
from which we deduce the expression that links the phase $\bar\gamma$ with the physical observables:
\bea
\tan\bar\gamma= \frac{ (m_2 -m_1)
\tan\frac{\hat\theta_e+\hat\theta_\mu}{2} + (m_1+m_2)
\tan{\phi \over 2}}{ (m_2 -m_1) -(m_1+m_2)
\tan\frac{\hat\theta_e+\hat\theta_\mu}{2}\,\tan{\phi \over 2}}\,.
\eea

Additional details of the connection between light neutrino
mass and flavor eigenstates, based on the
analysis of Ref.~\cite{Barroso}, are given in Appendix 2.

\subsubsection{Oscillation probabilities for $d\le5$-Effective Theory}

For two generations of leptons, the neutrino-neutrino
oscillation probability
is given explicitly by 
\bea 
\label{probab}
P(\nu_e\to\nu_\mu)&=&
\sin^2{2\theta}\,
\sin^2\left(\frac{
\Delta
m^2\,L}{4E}\right) = \left(1 - P(\nu_e \to \nu_e) \right),\eea
and the
neutrino-antineutrino
oscillation
probabilities are \cite{Majorana2}:
\bea
\label{probab1}
P(\nu_e\to\bar\nu_e)&=&
\frac{m_1^2}{E^2}\,\cos^2{\theta}+
\frac{m_2^2}{E^2}\,\sin^2{\theta}-P(\nu_e \to \bar\nu_\mu),\nn\\
P(\bar\nu_e\to\nu_e)&=&
\frac{m_1^2}{E^2}\,\cos^2{\theta}+
\frac{m_2^2}{E^2}\,\sin^2{\theta}-P(\bar\nu_e \to \nu_\mu),\nn\\
P(\nu_\mu\to\bar\nu_\mu)&=&
\frac{m_1^2}{E^2}\,\sin^2{\theta}+
\frac{m_2^2}{E^2}\,\cos^2{\theta}-P(\nu_\mu \to \bar\nu_e),\\
P(\bar\nu_\mu\to\nu_\mu)&=&
\frac{m_1^2}{E^2}\,\sin^2{\theta}+
\frac{m_2^2}{E^2}\,\cos^2{\theta}-P(\bar\nu_\mu \to \nu_e),\nn\\
P(\nu_e\to\bar\nu_\mu)&=&
\frac{\sin^2{2\theta}}{4E^2}
\left[m_1^2+m_2^2
-2m_1m_2\,\cos\left(\phi -\frac{ \Delta m^2\,L}{2E}\right)\right]
= P(\nu_\mu \to \bar\nu_e),\nn\\
P(\bar\nu_e\to\nu_\mu)&=& \frac{\sin^2{2\theta}}{4E^2}
\left[m_1^2+m_2^2
-2m_1m_2\,\cos\left(\phi +\frac{ \Delta m^2\,L}{2E}\right)\right]
=P(\bar\nu_\mu \to \nu_e),\nn
\eea
where $\Delta m^2 = m_2^2 - m_1^2$.

It was pointed out in  Ref.~\cite{gkm}  that the phase $\phi$ can  be 
extracted by measuring the asymmetry
\bea
{\cal A}&=& \frac{P(\nu_e\to\bar\nu_\mu)-P(\bar\nu_e\to\nu_\mu)}
{P(\nu_e\to\bar\nu_\mu)+P(\bar\nu_e\to\nu_\mu)}=
\frac{-2m_1m_2\,\sin\phi\,\sin\left(\frac{ \Delta m^2\,L}{2E}\right)}
{m_1^2+m_2^2-2m_1m_2\,\cos\phi\,\cos\left(\frac{ \Delta
     m^2\,L}{2E}\right)}~,
\eea
in flavor-changing neutrino oscillations.

Neutrinoless double beta ($0\nu 2\beta$) decay experiments are sensitive to the $CP$-even
quantity 
\bea
\label{0nu2beta}
\langle m_{ee}\rangle
=\left|m_1\,\cos^2\theta+m_2\,\sin^2\theta\,e^{-i\phi}\right|
\equiv \left|\sum_i  U_{ei}^2\, m_i \right|~,
\eea
which depends on the $CP$-odd phase $\phi$.  It is worth noting that the 
oscillation probability,
\bea
\label{pebe}
P(\nu_e\to\bar\nu_e)&=& \frac{1}{E^2}\,
 \left|m_1\,\cos^2{\theta}+m_2\,\sin^2{\theta} \,
e^{-i(\phi -\frac{ \Delta m^2\,L}{2E})}\right|^2\, ,
\eea
is proportional to a related combination with a different phase.

In the case of degenerate neutrinos, $P(\nu_e\to\bar\nu_e)$ reduces to
\bea
\label{pebedeg}
P(\nu_e\to\bar\nu_e)&=& \frac{m^2}{E^2}\,
 (\cos^4{\theta}\,+\,\sin^4{\theta} \,+\, 2\,\sin^2{\theta}
 \cos^2{\theta}\,\cos{\phi}) \ne \frac{m^2}{E^2},
\eea 
which shows that the mixing angle $\theta$ remains physical for 
degenerate Majorana neutrinos so long as 
$\phi \ne 0  $.
One recovers the naive expectation $m^2/E^2$ only for $\phi=0, \pi$.
A non-trivial dependence of flavor-changing neutrino-antineutrino
oscillation probabilities
on $\theta$ and $\phi$ in the degenerate limit was noticed in Ref.~\cite{gkm}.
\eq{pebedeg} shows that flavor-conserving transitions also depend on $\theta$
and $\phi$ in the degenerate limit.

\subsubsection{Low-energy ${d\le6}$-Effective Lagrangian}

Now consider the neutrino Lagrangian in \eq{d6_lagr} where the neutrino
kinetic 
energy term is no longer diagonal in flavor. In the physical low-energy basis,
the $2\times 2$ 
Hermitian matrix $\lambda$ of the light neutrinos depends on
3 real and 1 imaginary parameters
\bea 
\label{epsilon}
\lambda = {v^2 \over 2} c^{d=6} \equiv
\,\left(\begin{array}{cc}
\lambda_{ee}& \lambda_{e \mu}\,e^{i\bar\sigma} \\
\lambda_{e \mu}\,e^{-i\bar\sigma}& \lambda_{\mu\mu} \\
\end{array}\right)\, .
\eea

By performing the rephasing of the charged lepton fields given in
Eqs.~(\ref{reph_charg_lep}) and~(\ref{zeta_i}), the physical
weak currents can be described in terms of the mixing matrix
 $U^{\rm{eff}}$, \eq{Ueff},
\bea 
U^{\rm{eff}}\equiv
\left( 1 - \Omega \, {\lambda \over 2} \, \Omega^\dagger \right) \, U = 
\left(
\begin{array}{cc}
1- \frac{\lambda_{ee}}{2}& - \frac{\lambda_{e \mu}}{2}\,e^{i\Sigma} \\
- \frac{\lambda_{e \mu}}{2}\,e^{-i\Sigma}& 1- \frac{\lambda_{\mu\mu}}{2} \\
\end{array}
\right)\, U \, .
\label{U_eff}
\eea
The physical lepton mixing matrix $U^{\rm{eff}}$
depends on two phases: the phase $\phi$ of $U$ given in \eq{U}, 
and the phase
\bea
\label{Sigma}
\Sigma \equiv \bar\sigma + \left( \omega_e - \omega_\mu \right) =
\bar\sigma + \frac{\phi}{2}+ \frac{\hat\theta_{e} - \hat\theta_{\mu}}{2}\,=\,
\bar\sigma +\rho ~.
\eea
The low-energy phases $\phi$ and $\Sigma$ (or $\rho$ and $\bar\sigma$ in the 
parameterization in terms of $V^{\rm{eff}}$ ) are both
invariant under $V_\ell=V_e$ transformations, as previously stated.
 
Notice that the phase $\Sigma$ depends on the phases $\hat\theta_\alpha$,
so that these phases cannot be completely removed from the theory
when the $d=6$ operator is included.

\subsubsection {Oscillation probabilities for $d\le6$-Effective Theory}

The oscillation probabilities derived with the inclusion of only the $d=5$
Majorana mass term are modified by the inclusion of the $d=6$ matrix $\lambda$.
The neutrino-neutrino oscillation probabilities are given by 
\bea
\label{probab2}
P(\nu_e\to\nu_e)&=& (1 - 2\lambda_{ee} )-
\left[ (1 - 2\lambda_{ee} ) \sin^2 2\theta + 2 \,\lambda_{e \mu}\, \sin 2 \theta \,\cos2\theta
\,\cos(\Sigma) \right]\,
\sin^2\left(\frac{ \Delta m^2\,L}{4E}\right)~,\nn\\
P(\nu_\mu \to\nu_\mu)&=& (1 - 2\lambda_{\mu\mu} )-
\left[ (1 - 2\lambda_{\mu\mu} ) \sin^2 2\theta - 2 \,\lambda_{e \mu}\, \sin 2 \theta \,\cos2\theta
\,\cos(\Sigma) \right]\,
\sin^2\left(\frac{ \Delta m^2\,L}{4E}\right)~,\nn\\
P(\nu_e\to\nu_\mu)&=& \left[1-(\lambda_{ee}+\lambda_{\mu\mu})\right]\,\sin^2{2\theta}\,
\sin^2\left(\frac{ \Delta m^2\,L}{4E}\right)
+\lambda_{e \mu}\,\sin{2\theta}\,\sin(\Sigma)\,
\sin\left(\frac{ \Delta m^2\,L}{2E}\right)~, \nn\\
P(\nu_\mu\to\nu_e)&=& \left[1-(\lambda_{ee}+\lambda_{\mu\mu})\right]\,\sin^2{2\theta}\,
\sin^2\left(\frac{ \Delta m^2\,L}{4E}\right)
-\lambda_{e \mu}\,\sin{2\theta}\,\sin(\Sigma)\,
\sin\left(\frac{ \Delta m^2\,L}{2E}\right)~,\nn\\
\eea
where $\Delta m^2 = m_2^2 - m_1^2$.
With the inclusion of the $d=6$ operator, 
the sum of the $\nu_e \to \nu_\alpha $ oscillation probabilities no longer equals unity:
\bea
P(\nu_e\to\nu_e)+ P(\nu_e\to\nu_\mu)&=& (1 - 2\lambda_{ee}) + \left[
\left( \lambda_{ee} - \lambda_{\mu\mu}\right) 
\, \sin^2 2 \theta \,
- \lambda_{e \mu}\,\sin4 \theta \, \cos(\Sigma)\,\right]
\sin^2\left(\frac{ \Delta m^2\,L}{4E}\right)
\nn\\
&+& \lambda_{e \mu}\,\sin2 \theta\,  \sin(\Sigma)\,
\sin \left(\frac{ \Delta m^2\,L}{2E}\right)~.
\eea

The main consequence of the inclusion of ${d=6}$ operator is the
appearance of $CP$-violation in conventional neutrino-neutrino
oscillations with
only 2 neutrino families.
The $CP$-odd phase $\Sigma$ leads to the asymmetry
\bea
{\cal A}&=& \frac{P(\nu_e\to\nu_\mu)-P(\bar\nu_e\to\bar\nu_\mu)}
{P(\nu_e\to\nu_\mu)+P(\bar\nu_e\to\bar\nu_\mu)}=
\,4\,\lambda_{e \mu}\frac{\sin\Sigma}{\sin{2\theta}}\cot\left(\frac{ \Delta
     m^2\,L}{4E}\right)~,
\eea
where $CPT$ invariance has been used, i.e. $P(\nu_\mu\to\nu_e)
=P(\overline\nu_e\to\overline\nu_\mu)$.

The neutrino-antineutrino oscillations probabilities and $0\nu\beta\beta$
decay formulae also
are modified when the  
$d=6$ operator is included. The explicit formulae are lengthy and are
relegated to Appendix 3.

\subsection{Connection between High- and Low-Energy Parameters}

We can now apply the formulae obtained in Sect.~4 to the case of two heavy 
and two light neutrinos.

In the physical low-energy basis,
the $2 \times 2$ Hermitian matrix
\be
\lambda =\left(\begin{array}{cc}
\lambda_{ee}& \lambda_{e\mu} \, e^{i \bar\sigma} \\
\lambda_{e\mu} \, e^{-i \bar\sigma}& \lambda_{\mu \mu} \\
\end{array}\right)
\ee
is diagonalized by
the $2 \times 2$ unitary matrix $W$:
\be\label{epsdiag}
\lambda = W \, \lambda_{\rm diag, \, real} \, W^\dagger \,= 
W \, \left(\begin{array}{cc}
\lambda_1& 0 \\
0& \lambda_2 \\
\end{array}\right) \, W^\dagger \, , 
\ee
where $W$ is parameterized by 1 mixing angle $\theta^\prime$ and 
1 phase $\bar\sigma$,
\bea
W &=&
\left(\begin{array}{cc}
\cos\theta^\prime& \sin\theta^\prime \, e^{-i\bar\sigma} \\
-\sin\theta^\prime \, e^{i\bar\sigma}& \cos\theta^\prime \\
\end{array}\right)~.
\eea

The matrix $V_N$ is the unitary matrix which diagonalizes the heavy Majorana
neutrino mass matrix $M$,
\bea\label{BigMdiag}
M_{\rm diag, real} &\equiv& V_N^* \,{M}\,V_N^\dagger\, ,
\eea
where $M$ is given in terms of the low-energy parameters by
\bea
\label{M1}
{M}&\equiv&-(\sqrt{\lambda_{\rm diag, \, real}})^{-1} \, W^\dagger \,V^\dagger \,m_{\rm diag,\,real}\,V^* 
W^*\, (\sqrt{\lambda_{\rm diag, \, real}})^{-1} \, ,
\eea
where $V$ is given by Eq.~(\ref{V}).  Thus, $M$ depends on 
the low-energy phases $\hat\theta_e$, $\hat\theta_\mu$, $\phi$ and $\bar\sigma$,
the eigenvalues of the matrices $m$ and $\lambda$, and
the mixing angles $\theta$ and $\theta^\prime$.
$V_N$ can be determined as a function of these low-energy parameters as well. 

It is useful to rewrite the above in terms of the physical lepton mixing 
matrix $U$ defined by Eq.~(\ref{Omega}).
Define the matrix
\bea
\tilde W\equiv \Omega\,W\,\Omega^\dagger= 
\left(\begin{array}{cc}
\cos\theta^\prime& \sin\theta^\prime \, e^{-i\Sigma} \\
-\sin\theta^\prime \, e^{i\Sigma}& \cos\theta^\prime \\
\end{array}\right)\,,
\eea 
which diagonalizes
\be
\tilde\lambda\equiv \Omega\,\lambda\,\Omega^\dagger= 
\left(\begin{array}{cc}
\lambda_{ee}& \lambda_{e\mu} \, e^{i \Sigma} \\
\lambda_{e\mu} \, e^{-i \Sigma}& \lambda_{\mu \mu} \\
\end{array}\right)\,.
\ee

\eq{BigMdiag} can be rewritten as
\bea
M_{\rm diag,\,real} &=&{\tilde V_N}^* \,{\tilde M} \,{\tilde V_N}^\dagger\,,
\eea
where
\bea
\tilde V_N \equiv  V_N \, \Omega^T
\eea
is the unitary matrix which diagonalizes
\bea\label{Mtilde}
\tilde M&\equiv& \Omega \, M \, \Omega^T =
- (\sqrt{\lambda_{\rm diag, \, real}})^{-1} 
\, {\tilde W}^\dagger \,U 
\,m_{\rm diag,\,real}\,U^T\,
{\tilde W}^*\, (\sqrt{\lambda_{\rm diag, \, real}})^{-1} \, .
\eea
Notice that Eq.~(\ref{Mtilde}) implies that the imaginary part of $\tilde M$ 
depends on only the two physical phases $\phi$ and $\Sigma$.  
The explicit formulae for $\tilde M$ are given in Appendix 4.

Comparison of Eqs.~(\ref{M1}) and~(\ref{Mtilde}) with
Eq.~(\ref{M_L}) gives
\be\label{Ldef}
L = V^* \, W^* \, (\sqrt{\lambda_{\rm diag, \, real}})^{-1} \,  V_N^\dagger
= U^T \, \tilde W^* \, (\sqrt{\lambda_{\rm diag, \, real}})^{-1} \, {\tilde
V_N^\dagger}\, .
\ee
The matrix $L$ encoding the phase information relevant for 
leptogenesis is seen to depend on the phases $\phi$ and $\Sigma$.

The combination $Y_\nu^\dagger Y_\nu$ relevant for leptogenesis is given by
Eq.~(\ref{lepto_L}) with Eq.~(\ref{Ldef}) substituted for $L$,
\bea
Y_\nu^\dagger Y_\nu &=&{\tilde V_N} 
(\sqrt{\lambda_{\rm diag, \, real}})^{-1} \, {\tilde W}^* \,U^* 
\,m_{\rm diag,\,real}^2\,U^T\,
{\tilde W}^*\, (\sqrt{\lambda_{\rm diag, \, real}})^{-1}
\,{\tilde V_N}^\dagger \ .
\eea
The Majorana phases of the light neutrinos appearing in $D_{Maj}$ cancel in
the combination $U^* \,m_{\rm diag,\,real}^2\,U^T$. 
Thus, any dependence of the leptogenesis asymmetry
on the low-energy Majorana phases remains only through the $\tilde{V_N}$ matrix.

\section{Conclusions}

The low-energy 
$d\le6$-effective Lagrangian of the seesaw model is equal to 
the SM Lagrangian plus two higher-dimension operators: 
a $d=5$ operator, the well-known operator responsible for light Majorana neutrino masses, 
and a $d=6$ operator.
We have determined the connection between the high-energy Lagrangian of the seesaw model 
and the low-energy effective theory, when the number of heavy neutrinos is less than
or equal to the number of light lepton generations.

We have used the chiral symmetries of the seesaw Lagrangian and the effective theory
to define physical high-energy and low-energy bases which contain no unphysical parameters.
For practical purposes, a standard high-energy basis has been defined.  It does not 
exactly correspond to the usual physical low-energy basis, and the connection 
between the two has been discussed 
in detail.

We have illustrated how the measurement of low-energy neutrino-neutrino and neutrino-antineutrino transitions would allow {\it a priori} the determination 
of all the matrix elements of the $d=5$ and $d=6$ operator coefficients, and, thus, 
all the parameters of the high-energy seesaw Lagrangian. 
In particular, we have found a non-vanishing $CP$-asymmetry 
in neutrino-neutrino oscillations, which 
at leading order is sensitive to the $CP$-phases of the $d=6$ operator coefficient, and is 
non-vanishing even in the case of only two light neutrino species.

Although in practice the experimental determination of the $d=6$ operator is 
out of reach for natural values of the seesaw scale, the analysis here
 has strong 
implications for the 
explanation of the matter-antimatter asymmetry of the universe through leptogenesis.
The latter is shown to require both non-degenerate light neutrino masses and non-degenerate 
eigenvalues of the $d=6$ operator. When both of these degeneracies 
simultaneously occur, 
it is shown that the heavy neutrinos become mass degenerate and there is no leptogenesis
asymmetry.

Finally, the general results have been illustrated in full detail for the case of two 
light and two heavy neutrino generations.

\section{Acknowledgments}
We are indebted to Pilar Hernandez for discussions.  
A.B and M.B.G were partially
supported by CICYT FPA2000-0980 project.
A.B acknowledges MECD for financial support by
FPU grant AP2001-0521.
E.J. was supported in part by the Department
of Energy under grant DOE-FG03-97ER40546. 

\setcounter{section}{0}

\section{Appendix 1}

In a general basis, the complex symmetric $n^\prime \times n^\prime$
matrix $M$ contains $n^\prime(n^\prime+1)/2$ real and $n^\prime(n^\prime+1)/2$ imaginary parameters.  
Consider a basis in which $M$ is diagonal and complex, rather than diagonal
and real.  The transformation
\be
M \rightarrow V_N^* \, M \, V_N^\dagger = M_{\rm diag}=
{\rm diag}\left(|M_1|\,e^{i\theta_1}, |M_2|\,e^{i\theta_2}, \cdots, 
|M_{n^\prime}|\,e^{i\theta_{n^\prime}}\right)~, 
\ee
determines only $n^\prime(n^\prime-1)/2$ real and $n^\prime(n^\prime-1)/2$ imaginary parameters of 
the $n^\prime \times n^\prime$ unitary matrix $V_N$, which
contains $n^\prime(n^\prime-1)/2$ real and $n^\prime(n^\prime+1)/2$ imaginary parameters.  $M$ now depends 
on $n^\prime$ real and $n^\prime$ 
imaginary parameters.  $n^\prime$ imaginary parameters of $V_N$ are not determined
since $M_{\rm diag}$ remains diagonal under a transformation
\be\label{newV_N}
V_N^\dagger = {\rm diag}\left(e^{i\beta_1}, e^{i\beta_2}, \cdots,
e^{i\beta_{n^\prime}}\right)~,
\ee
which depends on $n^\prime$ arbitrary phases $\beta_i$.
This rephasing leaves the $n^\prime$ real parameters $|M_i|$ invariant,
but affects the heavy Majorana phases, 
\bea
\theta_i \rightarrow \theta_i + 2 \beta_i \ ,
\eea
as well as the phases $\phi_{\alpha j}^\prime$ of the matrix
\be 
\left( Y_\nu^\prime \right)_{{\alpha}j} \equiv u_{\alpha j}\, 
e^{i\phi^\prime_{\alpha j}}\ .
\ee 
Under $V_\ell=V_e$ and $V_N$ rephasings, $Y_\nu^\prime$ transforms as
\bea
\label{Ynu2}
Y_\nu^\prime &\rightarrow& V_\ell \,
Y_\nu^\prime \, V_N^\dagger\,,
\eea
where $V_\ell$ and $V_N^\dagger$ are given by Eqs.~(\ref{V_ell}) and~(\ref{newV_N}),
respectively, 
so that the phases $\phi^\prime_{\alpha j} \to \phi^\prime_{\alpha j} + \zeta_\alpha + \beta_j$
under the rephasing.

The heavy Majorana mass eigenstate neutrinos in this alternative 
basis are defined by
\be
N_i = e^{i\theta_i/2}\,N_{Ri} + e^{-i\theta_i/2}\,{N_{Ri}}^c \,\equiv 
\sqrt {\eta_i}\,N_{Ri} + \sqrt{{\eta_i}^*}\,{N_{Ri}}^c \ ,
\ee
and continue to satisfy the condition ${N_i}^c = N_i$.                                                                                                      

The above basis is related to the standard high-energy basis in the text by
a rephasing with $\beta_i=-\theta_i/2$ under which $M$ becomes diagonal and real, 
and $Y_\nu^\prime \to Y_\nu = Y_\nu^\prime \sqrt{\eta^*}$, with phases 
$\phi_{\alpha j} \equiv (\phi_{\alpha j}^\prime - \theta_j/2)$.  The phases 
$\phi_{\alpha j}$, which are used in the standard high-energy
basis, are invariant under $\beta_i$ rephasings.

The coefficients of the higher-dimensional operators of the low-energy 
effective theory do not depend on $\beta_i$ rephasings, so
the low-energy coefficients computed in a general basis are 
equal to those computed in the standard high-energy basis.

\section{Appendix 2}

Consider the low-energy theory including only the effects of the 
$d=5$ operator, for the case $n=n^\prime=2$ generations.
In the body of the paper, we expressed the flavor eigenstate parameters in terms of the 
mass eigenstate parameters. Here the inverse procedure \cite{Barroso} is shown in detail: 
the mass eigenstate parameters are given in terms of the flavor eigenstate parameters. 

The light neutrino mass matrix 
\be
\left(\begin{array}{cc}
m_{ee}\,e^{i\gamma_e} & m_{e \mu}\,e^{i\gamma_{e \mu}} \\
m_{e \mu}\,e^{i\gamma_{e \mu}}& m_{\mu\mu}\,e^{i\gamma_\mu}\\
\end{array}\right)
\ee
is brought to diagonal and real form 
with eigenvalues $m_1$ and $m_2$ by the matrix $V^\prime$
\bea 
\left(\begin{array}{cc}
m_1& 0 \\
0&m_2 \\
\end{array}\right)\, &=&
{V^\prime}^*\,
\left(\begin{array}{cc}
m_{ee}\,e^{i\gamma_e} & m_{e \mu}\,e^{i\gamma_{e \mu}} \\
m_{e \mu}\,e^{i\gamma_{e \mu}}& m_{\mu\mu}\,e^{i\gamma_\mu}\\
\end{array}\right)\,{V^\prime}^\dagger\,,
\label{diag_Barr}
\eea
where 
\bea\label{Vprime}
{V^\prime}^\dagger&=& \left(\begin{array}{cc}
e^{-i\gamma_e/2}& 0 \\
0& e^{-i\gamma_\mu/2} \\
\end{array}\right) \, V^\dagger
\eea
and $V$ is given by Eqs.~(\ref{V}) and (\ref{Vrho}).
\eq{diag_Barr} can be rewritten as Eq.~(\ref{22diag})
in terms of the basis-independent phase $\bar\gamma$ defined in
\eq{lowphasegam}.  Explicitly, 
\bea 
\left(\begin{array}{cc}
m_1 & 0 \\
0&m_2 \\
\end{array}\right) &=&
\left(\begin{array}{cc}
e^{-i\hat\theta_e/2}& 0 \\
0& e^{-i\hat\theta_\mu/2} \\
\end{array}\right)
\left(\begin{array}{cc}
\cos\theta&  -\sin\theta\,e^{i\rho} \\
\sin\theta\,e^{-i\rho}& \cos\theta \\
\end{array}\right) \nn\\
&\times&
\left(\begin{array}{cc}
m_{ee}\,& m_{e \mu}\,e^{i\bar\gamma} \\
m_{e \mu}\,e^{i\bar\gamma}& m_{\mu\mu}\\
\end{array}\right)\,\nn\\
&\times&
\left(\begin{array}{cc}
\cos\theta&  \sin\theta\,e^{-i\rho} \\
-\sin\theta\,e^{i\rho}& \cos\theta \\
\end{array}\right)\,
\left(\begin{array}{cc}
e^{-i\hat\theta_e/2}& 0 \\
0& e^{-i\hat\theta_\mu/2} \\
\end{array}\right)\,,
\label{diaglittlem}
\eea
where the parameterization of $V$ in Eq.~(\ref{Vrho}) has been used.

The correspondence between the results of Ref.~\cite{Barroso} and 
Eqs.~(\ref{m_ee})-(\ref{m_mumu}) then is obtained immediately by
bringing the light Majorana neutrino mass matrix to diagonal and
complex form with the rephasing 
\bea
V_{\ell}=
\left(\begin{array}{cc}
e^{-i\hat\theta_e/2}& 0 \\
0& e^{-i\hat\theta_\mu/2} \\
\end{array}\right)\,.
\eea

The diagonal entries in \eq{diaglittlem} yield
\bea
\label{eigenvalues}
m_1\,e^{i{\hat\theta_e}}&=& m_{ee}\,\cos^2{\theta}
- \,m_{e\mu}\,e^{i(\bar\gamma+\rho)}\,\sin{2\theta}
+ m_{\mu\mu}\,e^{i2\rho}\,\sin^2{\theta}\,,\nn\\
m_2\,e^{i{\hat\theta_\mu}}&=& m_{ee}\,e^{-i2\rho}\,\sin^2{\theta}
+\,m_{e\mu}\,e^{i(\bar\gamma-\rho)}\,\sin{2\theta}
+ m_{\mu\mu}\,\cos^2{\theta}\,.
\eea
The condition that the off-diagonal matrix elements vanish is \cite{Barroso}:
\be
\label{nondiagonal}
\left( m_{\mu\mu}\, e^{i\rho} - m_{ee} \, e^{-i\rho} \right) \sin 2 \theta =
2 \, m_{e\mu} \, e^{i \bar\gamma}\, \cos 2\theta \ .
\ee

Taking the real and the imaginary parts of \eq{eigenvalues} and
\eq{nondiagonal}, we obtain six conditions on the parameters
\bea
m_1\,\cos{{\hat\theta_e}}&=& m_{ee}\,\cos^2{\theta}
- \,m_{e\mu}\,\cos{(\bar\gamma+\rho)}\,\sin{2\theta}
+ m_{\mu\mu}\,\cos{2\rho}\,\sin^2{\theta}\,,
\label{Re_m1}
\\
m_2\,\cos{{\hat\theta_\mu}}&=& m_{ee}\,\cos{2\rho}\,\sin^2{\theta}
+\,m_{e\mu}\,\cos{(\bar\gamma-\rho)}\,\sin{2\theta}
+ m_{\mu\mu}\,\cos^2{\theta}\,,
\label{Re_m2}
\\
m_1\,\sin{{\hat\theta_e}}&=&
-\,m_{e\mu}\,\sin{(\bar\gamma+\rho)}\,\sin{2\theta}
+ m_{\mu\mu}\,\sin{2\rho}\,\sin^2{\theta}\,,
\label{Im_m1}
\\
m_2\,\sin{{\hat\theta_\mu}}&=& 
-m_{ee}\,\sin{2\rho}\,\sin^2{\theta}
+\,m_{e\mu}\,\sin{(\bar\gamma-\rho)}\,,
\sin{2\theta}
\label{Im_m2}
\\
0&=&-\left(m_{\mu\mu} - m_{ee} \right) \, \cos{\rho}\sin 2 \theta 
+ 2\, m_{e\mu} \, \cos{ \bar\gamma}\, \cos 2\theta \ ,
\label{Re_nondiag}
\\
0&=&-\left( m_{\mu\mu} + m_{ee} \right) \, \sin{\rho}\sin 2 \theta 
+2 \, m_{e\mu} \, \sin{\bar \gamma}\, \cos 2\theta \ ,
\label{Im_nondiag}
\eea

We can analyze these equations in different cases, three of them considered in Ref 
\cite{Barroso}:
\begin{itemize}
\item{Case 1: $\bar\gamma=0$ with $m_{ee} \ne m_{\mu\mu}$.}

In this case,  \eq{Im_nondiag} implies that $\rho=0,\pi$ and
\eq{Re_nondiag} yields
\be
\tan (2 \theta) = \frac{2 m_{e\mu}}{ m_{\mu\mu}-m_{ee}}\,.
\ee

Let's choose, for instance, $\rho=0$. \eq{Im_m1} and \eq{Im_m2} imply that
$\hat\theta_e=\hat\theta_\mu=0,\pi$, which means no $CP$ violation.
The same holds for $\rho=\pi$.

For instance, for $\rho=\hat\theta_e=\hat\theta_\mu=0$, the mass eigenvalues
read:
\bea
m_1 &=& m_{ee} \cos^2\theta + m_{\mu\mu} \sin^2\theta - 2 m_{e\mu} \sin\theta \cos\theta, \nn\\
m_2 &=& m_{ee} \sin^2\theta + m_{\mu\mu} \cos^2\theta + 2 m_{e\mu} \sin\theta \cos\theta,
\eea
and the mixing matrix appearing in \eq{U} is
\be
U=\left(\begin{array}{cc}
\cos\theta&  \sin\theta\, \\
-\sin\theta\,& \cos\theta \\
\end{array}\right)\ .
\ee

\item{Case 2a: $\bar\gamma \ne 0$ with $m_{ee} = m_{\mu\mu}$.}

Now, \eq{Re_nondiag} yields $\theta=\pi/4$. Substituting these values in
\eq{Im_nondiag}, one finds $\rho=0,\pi$.  Substituting this set of values
in \eq{Im_m1} and \eq{Im_m2} yields
\bea
m_1\,\sin{{\hat\theta_e}}&=& \mp m_{e\mu}\,\sin{\bar\gamma},
\label{Im_m1_2a}
\\
m_2\,\sin{{\hat\theta_\mu}}&=& \pm  m_{e\mu}\,\sin{\bar\gamma}.
\label{Im_m2_a2}
\eea
The real parts in \eq{Re_m1} and
\eq{Re_m2} result in the equations 
\bea
m_1\,\cos{{\hat\theta_e}}&=& m_{ee} \mp m_{e\mu}\,\cos{\bar\gamma},
\label{Re_m1_2a}
\\
m_2\,\cos{{\hat\theta_\mu}}&=& m_{ee} \pm  m_{e\mu}\,\cos{\bar\gamma}.
\label{Re_m2_2a}
\eea
Since $\bar\gamma \ne 0$, the $CP$ parities
$\hat\theta_e,\hat\theta_\mu\ne 0,\pi$.

Finally, the mixing matrix is given by
\be
U=\left(\begin{array}{cc}
\frac{1}{\sqrt{2}}& \frac{1}{\sqrt{2}} \\
-\frac{1}{\sqrt{2}}\,& \frac{1}{\sqrt{2}}\\
\end{array}\right)
\left(\begin{array}{cc}
\pm e^{-i\frac{\hat\theta_e-\hat\theta_\mu}{2}} & 0 \\
0& 1 \\
\end{array}\right)\,.
\ee

\item{Case 2b: $\bar\gamma \ne 0$ with $m_{ee} = -m_{\mu\mu}$.}

Now, \eq{Re_nondiag} yields $\theta=\pi/4$. Substituting these
values in \eq{Im_nondiag}, one finds $\rho=\pm\pi/2$.  \eq{Im_m1} and 
\eq{Im_m2} reduce to
\bea
m_1\,\sin{{\hat\theta_e}}&=&  \mp m_{e\mu}\,\cos{\bar\gamma},
\label{Im_m1_2b}
\\
m_2\,\sin{{\hat\theta_\mu}}&=& \mp  m_{e\mu}\,\cos{\bar\gamma}\ .
\label{Im_m2_2b}
\eea

The real parts in \eq{Re_m1} and \eq{Re_m2} are
\bea
m_1\,\cos{{\hat\theta_e}}&=& m_{ee} \pm m_{e\mu}\,\sin{\bar\gamma},
\label{Re_m1_2b}
\\
m_2\,\cos{{\hat\theta_\mu}}&=& -m_{ee} \pm  m_{e\mu}\,\sin{\bar\gamma},
\label{Re_m2_2b}
\eea
and the mixing matrix that governs the CC current is
\be
U=\left(\begin{array}{cc}
\frac{1}{\sqrt{2}}& \frac{1}{\sqrt{2}} \\
-\frac{1}{\sqrt{2}}\,& \frac{1}{\sqrt{2}}\\
\end{array}\right)
\left(\begin{array}{cc}
e^{i(\pm \frac{\pi}{2}-\frac{\hat\theta_e-\hat\theta_\mu}{2})}& 0 \\
0& 1 \\
\end{array}\right)\,.
\ee

Notice that the $CP$ parities can only be $\bar\theta_e=\bar\theta_\mu=0, \pi$ 
for the special case  $\bar\gamma = \pi /2 $ .

\item{Case 3: $\bar\gamma \ne 0$ with $m_{ee} \ne \pm m_{\mu\mu}$.}
In this case \eq{Re_nondiag} and \eq{Im_nondiag} can be rewritten as
\bea
\label{phiandtheta}
\tan\rho&=&
\frac{m_{\mu\mu}-m_{ee}}{m_{ee}+m_{\mu\mu}}\,\tan\bar\gamma~,\\
\tan 2\theta&=&
\frac{2\,m_{e\mu}}{m_{\mu\mu}-m_{ee}}\,
\frac{\cos\bar\gamma}{\cos\rho}~=\frac{2\,m_{e\mu}}{m_{ee}+m_{\mu\mu}}\,
\frac{\sin\bar\gamma}{\sin\rho}~,
\label{theta}
\eea
which, together with \eq{Re_m1}-\eq{Im_m2}, allow all physical low-energy
parameters to be determined. 

The physical mixing matrix is 
\be 
U=\left(\begin{array}{cc}
\cos\theta&  \sin\theta\, \\
-\sin\theta\,& \cos\theta \\
\end{array}\right)\left(\begin{array}{cc}
e^{i\phi/2}& 0 \\
0& 1 \\
\end{array}\right)\ ,
\ee
with $\phi$ given by \eq{rho}.
\end{itemize}

\section{Appendix 3}

For two generations, the neutrino-antineutrino oscillation probabilities 
including the $d=5$ and the dominant $d=6$ coefficient are given by
\bea
\label{probab3}
P(\nu_e\to\bar\nu_e)&=&(1-2\lambda_{ee})\,\frac{1}{E^2}
\left\{m_1^2\cos^4\theta+m_2^2\sin^4\theta
+m_1m_2\,\frac{\sin2\theta}{2}\cos\left(\phi -\frac{ \Delta m^2\,L}{2E}\right)\right\}\nn\\\nn\\
&+&\lambda_{e \mu}\frac{\sin{2\theta}}{E^2}\,
\left\{(m_1^2\cos^2\theta-m_2^2\sin^2\theta)\cos\Sigma \right.\\\nn\\
&+&\left.{m_1m_2} \left[\sin^2\theta 
\cos\left(\Sigma +\phi - \frac{ \Delta m^2\,L}{2E} \right) 
-\cos^2\theta \cos\left( \Sigma - \phi + \frac{ \Delta m^2\,L}{2E} \right) \right] \right\} \, ,\nn \\\nn\\\nn\\
P(\nu_\mu\to\bar\nu_\mu)&=&(1-2\lambda_{\mu\mu})\,\frac{1}{E^2}
\left\{m_1^2\sin^4\theta+m_2^2\cos^4\theta
+m_1m_2\,\,\frac{\sin2\theta}{2}\cos\left(\phi -\frac{ \Delta m^2\,L}{2E}\right)\right\}\nn\\\nn\\
&+&\lambda_{e \mu}\frac{\sin{2\theta}}{E^2}\,
\left\{(m_1^2\sin^2\theta-m_2^2\cos^2\theta)\cos\Sigma \right.\\\nn\\
&+&\left.{m_1m_2} \left[ \cos^2\theta
\cos\left(\Sigma -\phi + \frac{ \Delta m^2\,L}{2E} \right) 
-\sin^2\theta \cos\left( \Sigma + \phi - \frac{ \Delta m^2\,L}{2E} \right) \right] \right\} \, ,\nn \\\nn\\\nn\\
P(\nu_e\to\bar\nu_\mu)&=&
(1-\lambda_{ee}-\lambda_{\mu\mu})\,\frac{\sin^2{2\theta}}{4E^2}
\left[m_1^2+m_2^2
-2m_1m_2\,\cos\left(\phi -\frac{ \Delta m^2\,L}{2E}\right)\right]\nn\\
&-&\lambda_{e \mu}\,\frac{\sin{2\theta}}{2E^2}
\, \Delta m^2\,\cos\Sigma  = P(\nu_\mu \to \bar\nu_e) \, ,\nn\\\nn\\\nn\\
P(\bar\nu_e \to \nu_\mu) &=& 
P(\nu_e\to\bar\nu_\mu)\,\left[\phi\to -\phi \right]\,
=P(\bar\nu_\mu \to\nu_e) \, , \\\nn\\
P(\bar\nu_e\to\nu_e)&=& 
P(\nu_e\to\bar\nu_e)\,\left[\phi\to -\phi \right] \, ,\\\nn\\
P(\bar\nu_\mu\to\nu_\mu)&=& 
P(\nu_\mu\to\bar\nu_\mu)\,\left[\phi\to -\phi \right]\, .
\eea

The $CP$-even quantity probed in $0\nu\beta\beta$-decay becomes
\bea
\langle m_{ee}\rangle
&\equiv& \left| 
m_1\, \left(U_{e1}^{\rm{eff}}\right)^2 + m_2\, \left(U_{e2}^{\rm{eff}}\right)^2  \right| \\
&=&\left|\,m_1\,\cos^2\theta\, \left[ 1 - \lambda_{ee}\, +\lambda_{e \mu} \, \tan\theta \,
e^{i\Sigma} \right]
+m_2\,\sin^2\theta\,\left[ 1 - \lambda_{ee} - \lambda_{e \mu}
\cot\theta \, e^{i\Sigma}\right]\,e^{-i\phi} \right|.\nn
\eea

Comparison of this quantity with the oscillation
probability 
\bea
P(\nu_e\to\bar\nu_e)&=& \frac{1}{E^2}\,
\left| \,m_1\, \left(U_{e1}^{\rm{eff}}\right)^2 + m_2\, \left(U_{e2}^{\rm{eff}}\right)^2\,e^{i\,\frac{ \Delta m^2\,L}{2E}}  \right|^2\nn\\
&=&\frac{1}{E^2}\,\left|\,m_1\,\cos^2\theta \left[ 1 - \lambda_{ee}\, +\lambda_{e \mu} \, \tan\theta \,e^{i\Sigma} \right] \right.\\ç
&\,& \left.\qquad
+ m_2\,\sin^2\theta\,\left[ 1 - \lambda_{ee} - \lambda_{e \mu}
\cot\theta \, e^{i\Sigma}\right] \,
e^{-i\left(\phi-\frac{\Delta m^2\,L}{2E}\right)} \right|^2\nn\,,
\eea
shows that the simple relationship between these quantities
which was apparent when only
the $d=5$ operator is included (see Eqs.~(\ref{0nu2beta}) 
and~(\ref{pebe})) 
continues to hold when the $d=6$ operator is
included.

\section{Appendix 4}

In the case of two generations, the matrix $\tilde M$, defined in 
Eq.~(\ref{Mtilde}), has matrix elements
\bea
\tilde M
\equiv
\left(\begin{array}{cc}
{\tilde M}_{11}&{\tilde M}_{12}\\
{\tilde M}_{12}&{\tilde M}_{22}
\end{array}\right)~
=
\left(\begin{array}{cc}
|{\tilde M}_{11}|\,e^{i\,\Gamma_{11}}&|{\tilde M}_{12}|\,e^{i\,\Gamma_{12}}\\
|{\tilde M}_{12}|\,e^{i\,\Gamma_{12}}&|{\tilde M}_{22}|\,e^{i\,\Gamma_{22}}
\end{array}\right)~,
\eea
where the magnitudes $|{\tilde M}_{ij}|$ and 
phases $|\Gamma_{ij}|=\arg\{{\tilde M}_{ij}\}\,$ depend on the 
low-energy phases $\phi$ and $\Sigma$, the eigenvalues of $m$ and $\lambda$, 
and the mixing angles $\theta$ and $\theta^\prime$:
\bea
{\tilde M}_{11}&=&-\frac{e^{-i\Sigma}}{\lambda_1}
\left[\,m_1\,\,e^{i \phi}\,
\left(s^2\theta^\prime\,s^2\theta\,e^{-i \Sigma}+ c^2\theta^\prime\,c^2\theta\,e^{i \Sigma}+
\frac{1}{2}\, s2\theta^\prime\,s2\theta \right)\right.\nn\\
&\qquad& \qquad \;\;\; +\,\left.m_2\,\left( s^2\theta^\prime\,c^2\theta\,e^{-i\Sigma}
+c^2\theta^\prime\,s^2\theta\,e^{i\Sigma}
-\frac{1}{2}\, s2\theta^\prime\,s2\theta\,\right)\right]\,,\\
{\tilde M}_{22}&=&-\frac{e^{i\Sigma}}{\lambda_2}
\left[\,m_1\,\,e^{i\phi}\,
\left(s^2\theta^\prime\,c^2\theta\,e^{i\Sigma}
+c^2\theta^\prime\,s^2\theta\,e^{-i \Sigma} -
\frac{1}{2}\, s2\theta^\prime\,s2\theta\,\right)\right.\nn\\
&\qquad& \qquad \;\;\, +\,
\left.m_2\,\left(s^2\theta^\prime\,s^2\theta\,e^{i\Sigma}
+c^2\theta^\prime\,c^2\theta\,e^{-i\Sigma}
+\frac{1}{2}\, s2\theta^\prime\,s2\theta\,\right)\right]\,,\\
{\tilde M}_{12}&=&-\frac{1}{2\sqrt{\lambda_1 \lambda_2}}
\left[m_1\,e^{i\phi}
\left(s 2\theta^\prime\,(c^2\theta\,e^{i\Sigma}-s^2\theta\,e^{-i\Sigma})-
 c2\theta^\prime\,s2\theta\right)\right.\nn\\
&\qquad& \qquad \qquad\;\ +\,
\left.m_2 \left(s^2\theta^
\prime\,(s^2\theta\,e^{i \Sigma}- c^2\theta\,e^{-i\Sigma})+
\frac{1}{2}\, c2\theta^\prime\,s2\theta \right)\right]~,
\eea
where $s\equiv\sin$ and $c\equiv\cos$.

$\tilde{V_N}$ is parametrized in analogy to 
the diagonalizing matrix $V^\prime$ 
of the light Majorana neutrino mass matrix,
 Eqs.~(\ref{diag_Barr}) and~(\ref{V}),
\bea
\label{V_N}
{V_N}^\dagger &=&
\left(\begin{array}{cc}
e^{-i\Gamma_1/2}&0 \\
0& e^{-i\Gamma_2/2}\\
\end{array}\right)
\left(\begin{array}{cc}
\cos\Theta& \sin\Theta \, e^{-i\varrho} \\
-\sin\Theta \, e^{i\varrho}& \cos\Theta\\
\end{array}\right)
\left(\begin{array}{cc}
e^{-i\theta_1/2}&0 \\
0& e^{-i\theta_2/2}\\
\end{array}\right)\, ,
\eea
where $\Gamma_1 \equiv \Gamma_{11}$ and $\Gamma_2 \equiv \Gamma_{22}$.
\eq{Mdiag} then becomes
\bea
\left(\begin{array}{cc}
|M_1|&0\\
0&|M_2|
\end{array}\right)&=& 
\left(\begin{array}{cc}
e^{-i\theta_1/2}&0 \\
0& e^{-i\theta_2/2}\\
\end{array}\right)\,
\left(\begin{array}{cc}
\cos\Theta& -\sin\Theta \, e^{i\varrho} \\
\sin\Theta \, e^{-i\varrho}& \cos\Theta\\
\end{array}\right)
\nn\\\nn\\
&\times&\left(\begin{array}{cc}
|\tilde {M}_{11}^\prime| &|\tilde {M}_{12}^\prime|\,e^{i\,\bar\Gamma}\\
|\tilde {M}_{12}^\prime|\,e^{i\,\bar\Gamma}&|\tilde {M}_{22}^\prime|
\end{array}\right)\nn\\\nn\\
&\times&
\left(\begin{array}{cc}
\cos\Theta& \sin\Theta \, e^{-i\varrho} \\
-\sin\Theta \, e^{i\varrho}& \cos\Theta\\
\end{array}\right)\,\left(\begin{array}{cc}
e^{-i\theta_1/2}&0 \\
0& e^{-i\theta_2/2}\\
\end{array}\right)
~,
\eea
where 
\bea
\bar\Gamma \equiv \Gamma_{12}-\frac{\Gamma_{1}-\Gamma_{2}}{2}\,
\eea
depends on the low-energy real parameters and phases $\phi$ and $\Sigma$. 
Notice that this diagonalization is identical to that done for the light Majorana
neutrino mass matrix with the replacements 
\bea
\begin{array}{cc}
m_i \to|M_i|\,, & m_{\alpha\beta}\to |{\tilde M}_{ij}|\,,\nn\\\nn\\
\theta \to \Theta \,,&  \bar\gamma \to \bar\Gamma\,,\\\nn\\
\rho \to \varrho \,,& \hat\theta_\alpha \to \theta_i \,.\\
\end{array}
\eea

Following computations analogous to the ones performed in the diagonalization
of the light neutrino mass matrix, we can determine the heavy mass 
eigenvalues as a (complicated) function of the low-energy parameters.

\section{Appendix 5}

The Feynman rules of the high-energy seesaw model and the low-energy 
$d\le5$- and $d\le6$-effective theories are given in this appendix.

\subsection{Seesaw model}

\emph{\bf Yukawa couplings:}

In the basis in which $M$ is a diagonal real matrix, its mass
eigenstates are
\be
\label{field}
N_i \equiv \, {N_R}_i +  \, {N_{R\,i}}^c\, .
\ee
The Feynman rules for the Yukawa couplings of the heavy neutrinos to the light
lepton doublets are given by:

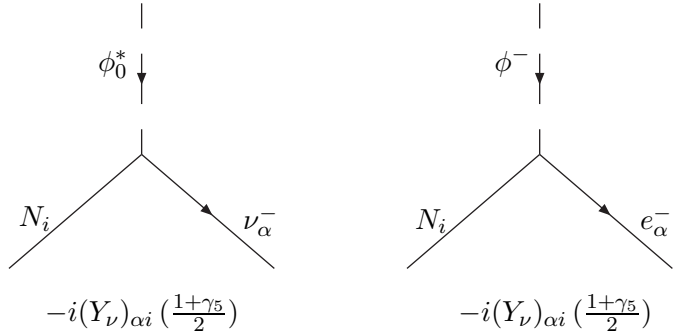
\begin{picture}(150,120)(-70,0)
\Line(0,0)(50,43)
\put(0,16){ $N_i$ }
\ArrowLine(50,43)(100,0)
\put(85,15){ $\nu_\alpha^-$}
\DashArrowLine(50,100)(50,43){10}
\put(30,75){ $\phi_{0}^*$}
\put(10,-20){ $-i(Y_{\nu})_{\alpha i}\,(\frac{1+\gamma_5}{2})$}
\Line(150,0)(200,43)
\put(150,15){ $N_i$ }
\ArrowLine(200,43)(250,0)
\put(235,15){ $e^-_\alpha$}
\DashArrowLine(200,100)(200,43){10}
\put(180,75){ $\phi^{-}$}
\put(170,-20){$-i(Y_\nu)_{\alpha i}\,(\frac{1+\gamma_5}{2})$}
\end{picture}
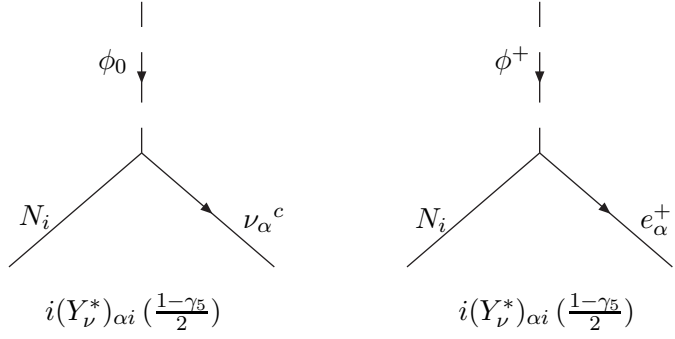
\vspace{1cm}

\begin{picture}(150,120)(-70,0)
\Line(0,0)(50,43)
\put(0,16){ $N_i$ }
\ArrowLine(50,43)(100,0)
\put(85,15){ ${\nu_\alpha}^c$}
\DashArrowLine(50,100)(50,43){10}
\put(30,75){ $\phi_{0}$}
\put(10,-20){ $i(Y_{\nu}^*)_{\alpha i}\,(\frac{1-\gamma_5}{2})$}
\Line(150,0)(200,43)
\put(150,15){ $N_i$ }
\ArrowLine(200,43)(250,0)
\put(235,15){ $e_\alpha^+$}
\DashArrowLine(200,100)(200,43){10}
\put(180,75){ $\phi^{+}$}
\put(170,-20){$i(Y_\nu^*)_{\alpha i}\,(\frac{1-\gamma_5}{2})$}
\end{picture}
\vspace{1cm}

\subsection{Effective theory}

\emph{\bf Charged currents: ${\cal L}_{SM}+\delta {\cal L}^{d=5}$}

In the basis in which $m$ is a diagonal real matrix and the mixing
matrix is
\be
U=\left(\begin{array}{cc}
\cos\theta&  \sin\theta\, \\
-\sin\theta\,& \cos\theta \\
\end{array}\right)\left(\begin{array}{cc}
e^{i\phi/2}& 0 \\
0& 1 \\
\end{array}\right)\, ,
\ee
the Feynman rules for the CC are:

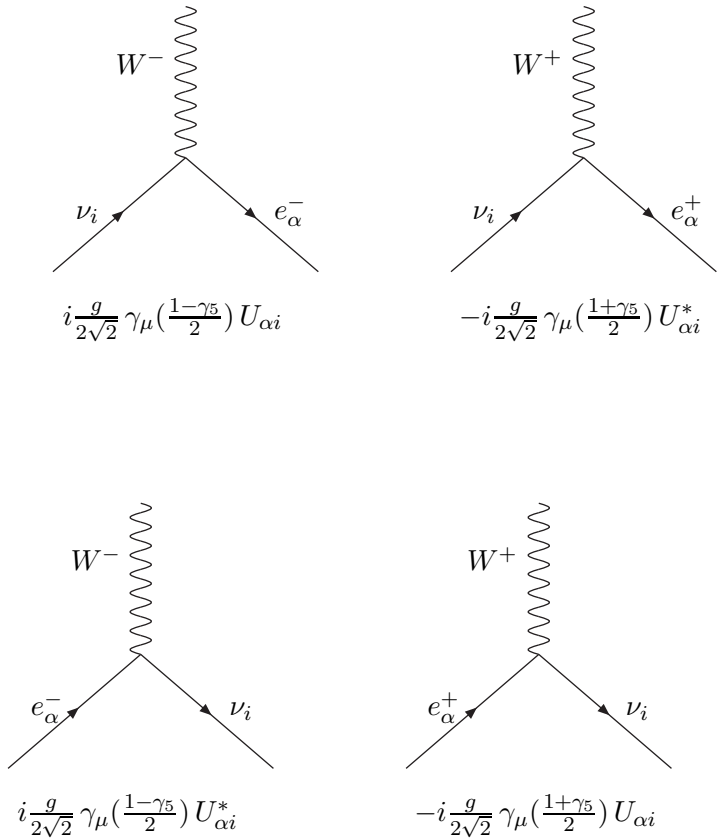
\begin{picture}(150,130)(-70,0)
\ArrowLine(0,0)(50,43)
\put(5,20){ $\nu_i$ }
\ArrowLine(50,43)(100,0)
\put(80,20){ $e_\alpha^-$}
\Photon(50,100)(50,43){4}{8}
\put(20,75){ $W^-$}
\put(0,-20){ $i\frac{g}{2\sqrt{2}}\, \gamma_\mu(\frac{1-\gamma_5}{2})\,
U_{\alpha i}$}
\ArrowLine(150,0)(200,43)
\put(155,20){ $\nu_i$ }
\ArrowLine(200,43)(250,0)
\put(230,20){ $e_\alpha^+$}
\Photon(200,100)(200,43){4}{8}
\put(170,75){ $W^{+}$}
\put(150,-20){ $-i\frac{g}{2\sqrt{2}}\, \gamma_\mu(\frac{1+\gamma_5}{2})\,
{U^*_{\alpha i}}$}
\end{picture}
\vspace{2cm}
\\
\begin{picture}(150,130)(-70,0)
\ArrowLine(0,0)(50,43)
\put(5,20){   $e_\alpha^-$}
\ArrowLine(50,43)(100,0)
\put(80,20){ $\nu_i$}
\Photon(50,100)(50,43){4}{8}
\put(20,75){ $W^-$}
\put(0,-20){ $i\frac{g}{2\sqrt{2}}\, \gamma_\mu(\frac{1-\gamma_5}{2})\,
U^*_{\alpha i} $}
\ArrowLine(150,0)(200,43)
\put(155,20){  $e_\alpha^+$}
\ArrowLine(200,43)(250,0)
\put(230,20){ $\nu_i$}
\Photon(200,100)(200,43){4}{8}
\put(170,75){ $W^{+}$}
\put(150,-20){ $-i\frac{g}{2\sqrt{2}}\, \gamma_\mu(\frac{1+\gamma_5}{2})\,
{U}_{\alpha i}$}
\end{picture}
\vspace{2cm}

\emph{\bf Charged currents: ${\cal L}_{SM}+\delta {\cal L}^{d=5}
+ \delta {\cal L}^{d=6}$}

In the same basis as before,
the Feynman rules for the charged currents, including both the $d=5$
and $d=6$ operators, are obtained from the above Feynman rules for the
$d\le5$-effective theory with the substitution
\be
U\to U^{\rm eff}
\equiv \left(\begin{array}{cc}
1- \frac{1}{2}\,
\lambda_{ee}& - \frac{1}{2}\,\lambda_{e \mu}\,e^{i\Sigma} \\
- \frac{1}{2}\,\lambda_{e \mu}\,e^{-i\Sigma}& 1- \frac{1}{2}\,\lambda_{\mu\mu} \\
\end{array}\right)\,U \, .
\ee


\end{document}